\documentclass[aps,prl,twocolumn,superscriptaddress]{revtex4-1}

\usepackage{graphicx,latexsym}
\usepackage{amsmath,amssymb,amsfonts}
\usepackage{bm}
\usepackage{braket}
\newcommand{\up}{\uparrow}
\newcommand{\down}{\downarrow}

\begin{document}


\title{$\eta$ Pairing of Light-Emitting Fermions:\\
Nonequilibrium Pairing Mechanism at High Temperatures
}
\author{Masaya Nakagawa}
\email{nakagawa@cat.phys.s.u-tokyo.ac.jp}
\affiliation{Department of Physics, University of Tokyo, 7-3-1 Hongo, Bunkyo-ku, Tokyo 113-0033, Japan}
\author{Naoto Tsuji}
\affiliation{Department of Physics, University of Tokyo, 7-3-1 Hongo, Bunkyo-ku, Tokyo 113-0033, Japan}
\affiliation{RIKEN Center for Emergent Matter Science (CEMS), Wako, Saitama 351-0198, Japan}
\author{Norio Kawakami}
\affiliation{Department of Physics, Kyoto University, Kyoto 606-8502, Japan}
\author{Masahito Ueda}
\affiliation{Department of Physics, University of Tokyo, 7-3-1 Hongo, Bunkyo-ku, Tokyo 113-0033, Japan}
\affiliation{RIKEN Center for Emergent Matter Science (CEMS), Wako, Saitama 351-0198, Japan}
\affiliation{Institute for Physics of Intelligence, University of Tokyo, 7-3-1 Hongo, Bunkyo-ku, Tokyo 113-0033, Japan}




\date{\today}

\begin{abstract}
Strongly interacting fermionic atoms are shown to develop $\eta$-pairing superfluid correlations in a nonequilibrium steady state in the presence of spontaneous emission of light from atoms. On the basis of the Hubbard model subject to spontaneous decay between internal spin states, we show that prohibition of radiative decay due to the Pauli exclusion principle and destructive interference between doublon-decay processes lead to nonequilibrium $\eta$ pairing. Because of the non-thermal nature of the steady state, pair correlations arise even from a completely uncorrelated infinite-temperature initial state, allowing coherent atom pairs to be formed at high temperatures. Experimental implementation with fermionic atoms in an optical lattice is discussed.
\end{abstract}


\maketitle

In a seminal paper \cite{Yang89}, Yang showed that exact eigenstates of the Fermi-Hubbard model can be constructed through the $\eta$-pairing mechanism. Although the Hubbard model in two and three dimensions is not exactly solvable due to the competing nature of hopping and interaction, a hidden $\eta$ symmetry of the model permits $\eta$-pairing eigenstates \cite{YangZhang90, Pernici90}. The $\eta$-pairing states exhibit an off-diagonal long-range order \cite{Yang62}, giving a rare example of fermionic superfluidity that is an exact eigenstate. However, since they cannot be the ground state of the Hubbard model except for the limiting case of an infinite attractive interaction \cite{Yang89, Singh91}, $\eta$ pairing requires nonequilibrium situations such as adiabatic state preparation \cite{Rosch08, Kantian10}, laser irradiation \cite{Kaneko19, Kaneko20, Werner19, Li20}, periodic driving \cite{Kitamura16, Peronaci20, Cook20, Tindall21, Tindall21_2}, and tailored dissipative processes \cite{Diehl08, Kraus08, Bernier13, Tindall19, XZZhang20, XZZhang21}.

Control of superfluidity with light lies at the forefront of nonequilibrium quantum many-body physics \cite{Fausti11, Kaiser14, Hu14, Mitrano16, Cantaluppi18, Suzuki19, Buzzi20, Budden20}. 
Since irradiation of light generally heats up a sample, how to create quantum coherence of a superfluid by the light-matter coupling is a highly nontrivial question with several interesting possibilities \cite{Kaneko19, Kaneko20, Werner19, Li20, Fausti11, Kaiser14, Hu14, Mitrano16, Cantaluppi18, Suzuki19, Buzzi20, Budden20}. 
Importantly, a pairing mechanism in nonequilibrium quantum matter may be fundamentally different from that of the BCS theory \cite{BCS57} because of its nonequilibrium nature.

\begin{figure}
\includegraphics[width=8.5cm]{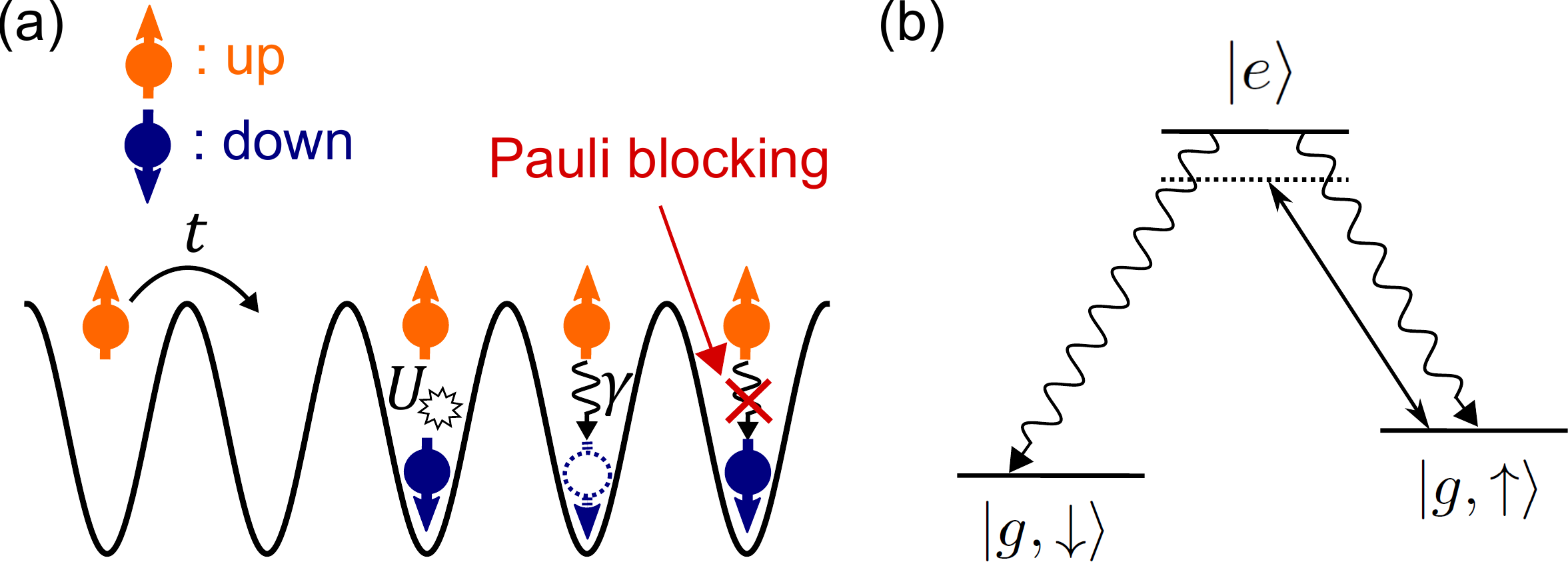}
\caption{(a) Schematic illustration of the model. Spin-$1/2$ fermionic atoms in an optical lattice are described by the Hubbard model with the hopping amplitude $t$ and the on-site interaction $U$, where spontaneous emission of light leads to an on-site decay process from spin-up to spin-down states with rate $\gamma$. When two atoms with opposite spins occupy the same site, the Pauli blocking prevents atoms from radiative decay. 
(b) Lambda-type level structure for the implementation of spontaneous decay in the Hubbard model. The spin-up state $\ket{g,\up}$ can decay to the spin-down state $\ket{g,\down}$ through an off-resonant coherent transition (shown by a double arrow) to an excited state $\ket{e}$ and a subsequent spontaneous decay (shown by wavy arrows) from it.}
\label{fig_setup}
\end{figure}

In this Letter, we propose a nonequilibrium $\eta$-pairing mechanism induced by spontaneous emission of light from strongly correlated fermionic atoms. 
We consider a cold-atom Fermi-Hubbard system in an optical lattice in which a spin-up state of an atom can decay to a spin-down state via a coupling to an excited state that decays by spontaneously emitting a photon, as illustrated in Fig.~\ref{fig_setup}. 
We show that the nonequilibrium steady states of this light-matter-coupled system are given by the $\eta$-pairing states. 
Here, the $\eta$-pairing states are stabilized due to double protection of fermion pairs from possible decay processes. First, as shown in Fig.~\ref{fig_setup}(a), the Pauli exclusion principle prohibits on-site pairs (i.e., doublons) from radiative decay \cite{Sandner11}. Second, complete destructive interference between doublon-decay processes prevents doublons from breaking up into single particles, thereby maintaining the $\eta$-pairing states as a coherent steady state. 
This pairing mechanism is inherently of nonequilibrium nature and makes a marked contrast to the BCS mechanism \cite{BCS57} in which an attractive interaction between fermions and the Fermi-surface instability play essential roles. 
Remarkably, unlike previous proposals \cite{Bernier13, Tindall19}, the $\eta$-pairing steady state is observed even when the dynamics starts from an infinite-temperature initial state, because the entropy of the system can be decreased by dissipative processes of spontaneous emission which is described by non-Hermitian Lindblad operators. 
This indicates that one can create coherent atom pairs at an initial temperature far above the BCS critical temperature. 
We propose an experimental implementation of the model with alkaline-earth-like fermionic atoms in an optical lattice and discuss how to detect the $\eta$-pairing state through measurement of a momentum distribution of doublons.


\textit{Model}.--\ 
We generalize a quantum optical master equation \cite{BreuerPetruccione}, which has been used to describe the dynamics of noninteracting atoms subject to spontaneous emission of photons, to interacting fermionic atoms \cite{Lehmberg70, Pichler10, Sarkar14}. We consider spin-$1/2$ fermionic atoms in a $d$-dimensional cubic optical lattice. The Hamiltonian of our system is described by the Hubbard model $H$ when the lattice potential is sufficiently deep and the single-band approximation is valid \cite{Esslinger10}:
\begin{align}
H=&-t\sum_{\langle i,j\rangle}\sum_{\sigma=\up, \down}(c_{i\sigma}^\dag c_{j\sigma}+\mathrm{H.c.})+U\sum_jn_{j\up}n_{j\down}\notag\\
&+\delta\sum_j(n_{j\up}-n_{j\down}),
\label{eq_Hubbard}
\end{align}
where $c_{j\sigma}$ is the annihilation operator of a spin-$\sigma$ fermion at a lattice position $\bm{R}_j\in\{0,\cdots,L-1\}^d$, $n_{j\sigma}\equiv c_{j\sigma}^\dag c_{j\sigma}$ is the density operator, $t$ is the tunneling amplitude, and $U$ is the on-site interaction strength. Here $\delta$ is the energy difference between the internal spin states, which can be set to zero without loss of generality \cite{supple}. 
We assume that atoms tunnel only between nearest-neighbor sites. 
We impose the periodic boundary condition to ensure the translational invariance of the model, but the following results can be generalized to the case of the open boundary condition. 

We consider a situation in which an atom in a spin-up state can decay to a spin-down state via spontaneous emission of a photon [see Fig.~\ref{fig_setup}(a)]. 
Such a decay process can be induced by weakly hybridizing the spin-up state with an excited level of the atom [see Fig.~\ref{fig_setup}(b)]. A detailed discussion will be given later and in Supplemental Material \cite{supple} with a concrete experimental implementation. 
Within the Born-Markov approximation, we can trace out the photon degrees of freedom and obtain the Lindblad-type quantum master equation that describes the time evolution of the density matrix $\rho$ of the atomic gas as \cite{BreuerPetruccione}
\begin{equation}
\frac{d\rho}{d\tau}=\mathcal{L}\rho=-i[H,\rho]+\sum_j(L_j\rho L_j^\dag-\frac{1}{2}\{ L_j^\dag L_j,\rho\}),
\label{eq_master}
\end{equation}
where $\tau$ is the time, $\mathcal{L}$ is the Liouvillian superoperator, and the Lindblad operator $L_j=\sqrt{\gamma}c_{j\down}^\dag c_{j\up}$ describes a spontaneous emission process with rate $\gamma>0$ at site $j$.


$\eta$-\textit{pairing steady state}.--\ 
The Hubbard Hamiltonian \eqref{eq_Hubbard} possesses the $\eta$ SU(2) symmetry generated by $\eta^x=(\eta^++\eta^-)/2,\eta^y=(\eta^+-\eta^-)/2i$, and $\eta^z=\frac{1}{2}\sum_j(n_{j\up}+n_{j\down}-1)$, where $\eta^+=\sum_j e^{i\bm{Q}\cdot\bm{R}_j}c_{j\up}^\dag c_{j\down}^\dag$ creates a doublon with crystal momentum $\bm{Q}=(\pi,\cdots,\pi)$, and $\eta^-=(\eta^+)^\dag$ \cite{Yang89, YangZhang90}. 
Thus, if a state $\ket{\Psi}$ is an eigenstate of $H$, so is $\eta^+\ket{\Psi}$. For example, Yang's $\eta$-pairing state $\ket{\psi_N}\equiv (\eta^+)^{N/2}\ket{0}$, where $\ket{0}$ is the vacuum of fermions, is an $N$-particle eigenstate of $H$ \cite{Yang89}. The $\eta$-pairing state can be regarded as a condensate of doublons at momentum $\bm{Q}$ and exhibits an off-diagonal long-range order \cite{Yang89, Yang62}.  
We also introduce a family of eigenstates $\ket{\psi_{2N_p};\bm{k}_1,\cdots,\bm{k}_{N_{\mathrm{ex}}}}\equiv(\eta^+)^{N_p}c_{\bm{k}_1\down}^\dag \cdots c_{\bm{k}_{N_{\mathrm{ex}}}\down}^\dag\ket{0}$, where $c_{\bm{k}\sigma}=\frac{1}{\sqrt{N_s}}\sum_j c_{j\sigma}e^{-i\bm{k}\cdot\bm{R}_j}$ is the annihilation operator of a single-particle eigenstate with crystal momentum $\bm{k}$, and $N_s$ denotes the number of lattice sites. This state also shows the $\eta$-pairing correlations but contains excess spin-down particles that do not form pairs.

Since $L_j\ket{\psi}=0$ for $\ket{\psi}=\ket{\psi_N}$ and $\ket{\psi}=\ket{\psi_{2N_p};\bm{k}_1,\cdots,\bm{k}_{N_{\mathrm{ex}}}}$, we find that the $\eta$-pairing states $\rho=\ket{\psi}\bra{\psi}$ are steady states of the quantum master equation \eqref{eq_master}: $\mathcal{L}\ket{\psi}\bra{\psi}=0$. 
Equivalently, no photon is emitted from the $\eta$-pairing states, implying that they are many-body dark states \cite{Diehl08, Kraus08}. 
The steadiness of the $\eta$-pairing states is physically understood as follows. First, on-site fermion pairs (doublons) are stabilized since a radiative decay of a spin-up particle is Pauli-blocked by a spin-down particle sitting at the same site [see Fig.~\ref{fig_setup}(a)] \cite{Sandner11}. Second, although doublons may have varying center-of-mass momenta and decay via hopping processes, $\eta$ pairs never decay since they constitute eigenstates of the Hubbard Hamiltonian. Physically, this is understood as complete destructive interference between different decay processes of doublons, as can be seen from $-t\sum_{\langle i,j\rangle}\sum_\sigma (c_{i\sigma}^\dag c_{j\sigma}+\mathrm{H.c.})\ket{\psi_N}=0$ \cite{supple}. 
The above pairing mechanism is clearly distinct from the BCS mechanism which arises from the Fermi-surface instability due to an attractive interaction between fermions.


\textit{Symmetries, conserved quantity, and exact Liouvillian eigenmodes}.--\ 
Since the system has many dark states, a general steady state is a statistical mixture of the $\eta$-pairing states $\ket{\psi_N}$ and $\ket{\psi_{2N_p};\bm{k}_1,\cdots,\bm{k}_{N_{\mathrm{ex}}}}$. 
The steady-state distribution of the $\eta$-pairing states depends on the initial state and the model parameters. 
To seek an optimal condition, we exploit the fact that the Lindblad operators commute with the generators of the $\eta$ SU(2) symmetry: $[L_j,\eta^+]=[L_j,\eta^-]=0$. This property and the $\eta$ SU(2) symmetry of the Hamiltonian lead to the conservation of
\begin{align}
C\equiv&\langle \eta^+\eta^-\rangle\notag\\
=&\sum_j\langle n_{j\up}n_{j\down}\rangle+\sum_{i,j\ (i\neq j)}e^{i\bm{Q}\cdot(\bm{R}_i-\bm{R}_j)}\langle c_{i\up}^\dag c_{i\down}^\dag c_{j\down}c_{j\up}\rangle,
\label{eq_conserv}
\end{align}
where $\langle \mathcal{O}\rangle=\mathrm{Tr}[\mathcal{O}\rho]$ \cite{Buca12, Albert14, Tindall19}. Thus, to achieve large $\eta$-pairing correlations in a steady state, we should prepare an initial state having sufficiently large $C$. 
This implies that high-energy or \textit{high-temperature} states rather than low-energy ones of the repulsive Hubbard model are appropriate for an initial state, since the former involve a large number of doublons which give a large $\sum_j\langle n_{j\up}n_{j\down}\rangle$ in Eq.~\eqref{eq_conserv}, whereas the latter do not. 
The conservation of $C$ is similar to the situation in Ref.~\cite{Tindall19}; however, we emphasize that the physical mechanism of the $\eta$ pairing is directly linked to the Fermi statistics and the destructive interference, but not to heating discussed in Ref.~\cite{Tindall19}.

The time scale of the formation dynamics of the $\eta$-pairing state is governed by the eigenvalues of the Liouvillian $\mathcal{L}$. 
Remarkably, some eigenvalues and the corresponding eigenmodes of the Liouvillian can exactly be obtained from the procedure in Refs.~\cite{Torres14, Nakagawa20}. 
To see this, we decompose the Liouvillian as $\mathcal{L}=\mathcal{K}+\mathcal{J}$, where $\mathcal{K}\rho\equiv -i(H_{\mathrm{eff}}\rho-\rho H_{\mathrm{eff}}^\dag)$ describes an effective non-Hermitian Hamiltonian dynamics with $H_{\mathrm{eff}}\equiv H-(i/2)\sum_j L_j^\dag L_j$, and $\mathcal{J}\rho\equiv \sum_j L_j\rho L_j^\dag$ is the quantum-jump term. 
The non-Hermitian Hamiltonian $H_{\mathrm{eff}}$ commutes with the total magnetization operator $S_z=\sum_j(n_{j\up}-n_{j\down})/2$, while the quantum-jump term decreases the magnetization by one. 
It follows that in the basis that diagonalizes $\mathcal{K}$ the Liouvillian $\mathcal{L}$ is a triangular-matrix form in which the off-diagonal elements are determined from $\mathcal{J}$. 
Thus, any Liouvillian eigenvalue $\lambda$ is the same as that of $\mathcal{K}$ and can be calculated from a pair of eigenvalues $E_a$ and $E_b$ of the non-Hermitian Hubbard model $H_{\mathrm{eff}}$ as $\lambda=-i(E_a-E_b^*)$ \cite{supple}. 
Such a feature does not exist in models with other Lindblad operators used in Refs.~\cite{Diehl08, Kraus08, Bernier13, Tindall19}. In particular, for the one-dimensional ($d=1$) case, the present model is exactly solvable by the Bethe ansatz method \cite{Nakagawa20, 1dHubbard_book}. 

The Liouvillian excitation spectrum of the $\eta$-pairing steady states involves single-particle and collective excitations as in conventional BCS superfluids. 
For example, single-particle excitations used to show the metastability of the $\eta$-pairing state \cite{Yang89} provide exact Liouvillian eigenmodes with degenerate eigenvalues $\lambda=-\gamma$ in any spatial dimensions \cite{supple}. 
Such single-particle modes decay on the time scale of $1/\gamma$. The late-stage dynamics near the steady state is governed by collective modes of the $\eta$ SU(2) pseudospins. In fact, the Bethe ansatz solution shows the dispersion relation of the collective modes as $\lambda(P)=2\mathrm{Im}\sqrt{16t^2\cos^2(P/2)+(U+i\gamma/2)^2}-\gamma\simeq 2\mathrm{Im}[J](1+\cos P)$ for $P\simeq \pi$, where $P$ is the momentum of the excitation and $J=4t^2/(U+i\gamma/2)$ is an effective exchange coupling between pseudospins, giving a gapless Liouvillian spectrum that implies a power-law tail in the relaxation dynamics in the thermodynamic limit \cite{Cai13, supple}. 
Note that the dispersion relation of the collective modes near $P=\pi$ is quadratic rather than linear in contrast to the Nambu-Goldstone mode of a conventional charge-neutral superfluid. This property highlights the difference between $\eta$-pairing and BCS superfluids; the relevant symmetry of the former is SU(2), while that of the latter is U(1).

Moreover, by the Shiba transformation \cite{Shiba72} $Vc_{j\up}V^\dag = c_{j\up}, Vc_{j\down}V^\dag =e^{i\bm{Q}\cdot\bm{R}_j}c_{j\down}^\dag$, the model discussed in this Letter can be mapped to a dissipative Hubbard model subject to two-body particle loss \cite{Nakagawa19, Nakagawa20} described by a Lindblad operator $e^{i\bm{Q}\cdot\bm{R}_j}V L_j V^\dag= \sqrt{\gamma}c_{j\down}c_{j\up}$. Thus, the Liouvillian spectrum of the model described in Eqs.~\eqref{eq_Hubbard} and \eqref{eq_master} is equivalent to that of a dissipative Hubbard model subject to two-body particle loss after changing the sign of the interaction strength. 
The $\eta$-pairing states of the model correspond to ferromagnetic steady states discussed in Refs.~\cite{Nakagawa19, Nakagawa20} through this mapping, while the physical mechanisms of their stabilization are different. 
This correspondence extends the Shiba symmetry between the repulsive and attractive Hubbard models to the symmetry between spontaneous emission and particle loss, making a nontrivial connection between different dissipative processes of many-body systems.


\begin{figure}
\includegraphics[width=8.5cm]{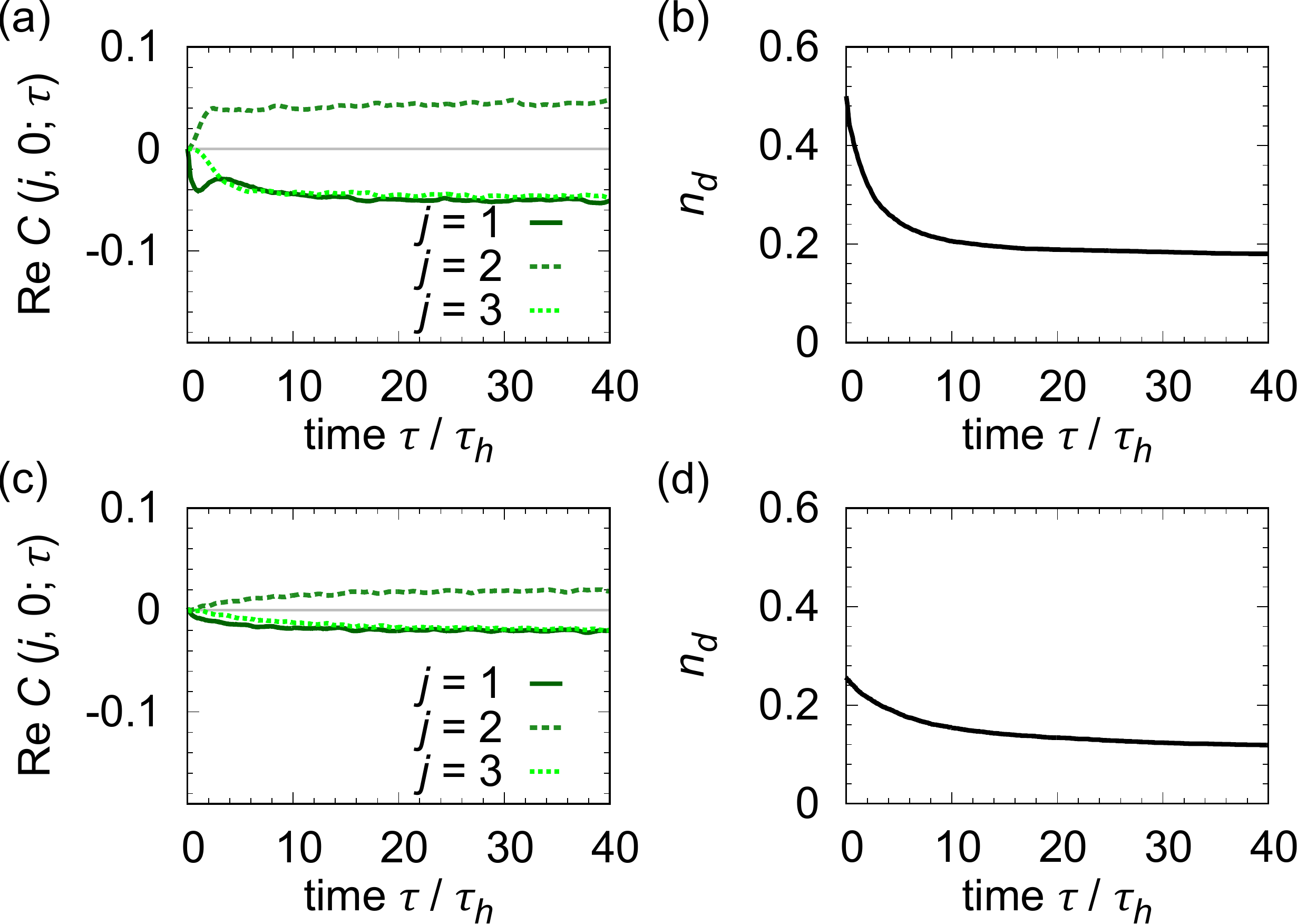}
\caption{Dynamics of the pair correlation function $\mathrm{Re}[C(j,0;\tau)]$ [(a), (c)] and the density $n_d$ of doublons [(b), (d)] for the one-dimensional system with 8 sites and 8 particles. The initial state is the doublon-density-wave state for (a) and (b), and the infinite-temperature state for (c) and (d). The model parameters are set to $t=1, U=10, \delta=0$, and $\gamma=10$. The unit of time is $\tau_h=1/t$.}
\label{fig_paircorr}
\end{figure}

\textit{Dynamics of pair correlations}.--\ 
We numerically solve the quantum master equation \eqref{eq_master} in the one-dimensional case by the quantum-trajectory method \cite{Dalibard92, Carmichael_book, Dum92, Daley14}. We take 1000 quantum trajectories, where the time evolution is calculated by exact diagonalization of the non-Hermitian Hamiltonian $H_{\mathrm{eff}}$. 
We consider an initial state $\rho(0)=\ket{\psi_0}\bra{\psi_0}$ of an 8-site system, where $\ket{\psi_0}=\prod_{j=0,2,4,6}c_{j\up}^\dag c_{j\down}^\dag\ket{0}$ is a density-wave state of doublons. 
Figure \ref{fig_paircorr} (a) shows the dynamics of the pair correlation function $C(j,0;\tau)\equiv\mathrm{Tr}[c_{j\up}^\dag c_{j\down}^\dag c_{0\down}c_{0\up}\rho(\tau)]$. In the steady state, the amplitude of the pair correlation function does not depend on sites and its sign alternates from site to site, indicating the formation of $\eta$ pairs with the center-of-mass momentum $\bm{Q}$. The imaginary part of the pair correlation function is negligible in the steady state, while the real part does not significantly depend on the model parameters \cite{supple}. 
Figure \ref{fig_paircorr}(b) shows the dynamics of the density of doublons $n_d\equiv(1/N_s)\sum_j\langle n_{j\up}n_{j\down}\rangle$. After a long time, the density of doublons reaches a nonzero steady-state value. 
Since the number of doublons in the initial state is $\mathcal{O}(N_s)$ and $C\geq 0$, the amplitude of the pair correlations in the steady states is estimated to be $\mathcal{O}(1/N_s)$, which can be observed as a long-range order in finite-size systems such as trapped atoms \cite{Tindall21, Mazurenko17}. 
The dynamics in Fig.~\ref{fig_paircorr} exhibits a two-step relaxation. Since the decay rate $\gamma$ is sufficiently larger than the hopping amplitude, the pair correlations initially increase over the time scale of hopping which breaks the doublons in the initial state, and the slow relaxation at a late stage near the steady state is governed by the collective modes with an exchange coupling $\mathrm{Im}[J]=-0.16$.

We also study the dynamics from an infinite-temperature initial state $\rho(0)=I/d_{\mathrm{H}}$, where $I$ and $d_{\mathrm{H}}$ are the identity operator and the dimension of the Hilbert space of the 8-particle sector, respectively. To numerically simulate the dynamics with the quantum-trajectory method, we sample a random pure initial state in each quantum trajectory. Figures \ref{fig_paircorr} (c) and (d) show the dynamics of the pair correlations and that of the density of doublons, respectively. Notably, the $\eta$-pairing correlations develop even if the dynamics starts from the infinite-temperature initial state. This behavior is in marked contrast to the heating-induced $\eta$-pairing mechanism in Ref.~\cite{Tindall19}, where the $\eta$ paring cannot develop from the infinite-temperature state because of the Hermiticity of Lindblad operators.

\begin{figure}
\includegraphics[width=8.5cm]{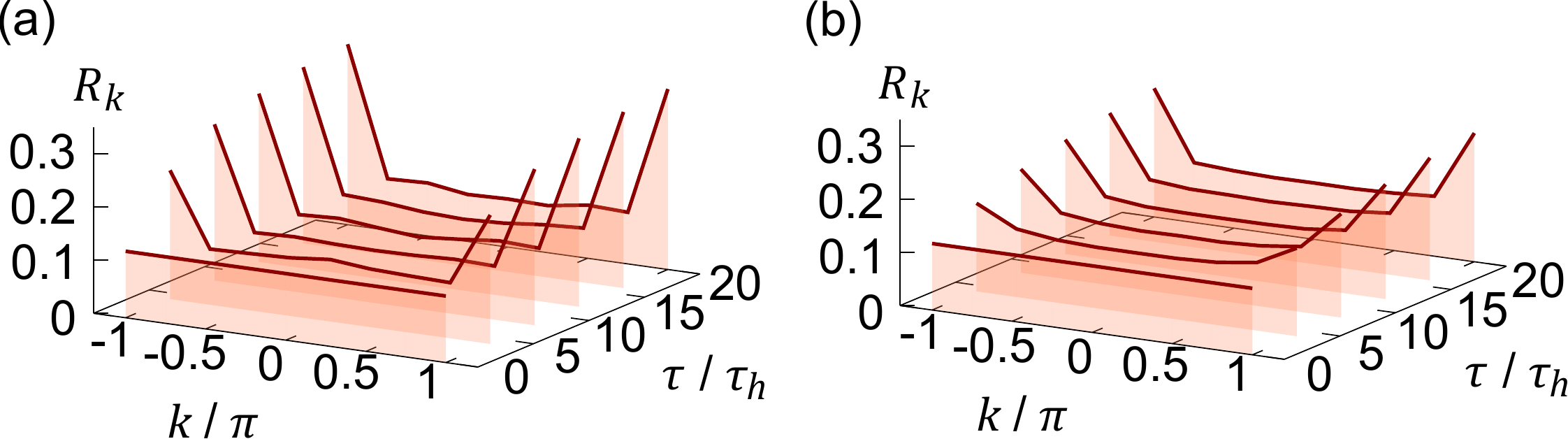}
\caption{Dynamics of the normalized momentum distribution $R_k$ of doublons for (a) the doublon-density-wave initial state and (b) the infinite-temperature initial state. The model parameters are the same as those in Fig.~\ref{fig_paircorr}. The unit of time is $\tau_h=1/t$.}
\label{fig_momdist}
\end{figure}

\textit{Doublon momentum distribution}.--\ 
The formation of an $\eta$-pairing state can be detected from the momentum distribution of doublons, which can be measured with time-of-flight techniques after doublons are converted into molecules by a Feshbach resonance \cite{Regal04, Zwierlein04, Stoferle06, Altman05, Perali05, Matyjaskiewicz08}. The momentum distribution of doublons is defined as $R_{\bm{k}}\equiv\langle d_{\bm{k}}^\dag d_{\bm{k}}\rangle/\sum_{\bm{k}}\langle d_{\bm{k}}^\dag d_{\bm{k}}\rangle$, where $d_{\bm{k}}\equiv (1/\sqrt{N_s})\sum_j c_{j\down}c_{j\up}e^{-i\bm{k}\cdot\bm{R}_j}$ is the annihilation operator of a doublon with momentum $\bm{k}$. Here, we have normalized the momentum distribution by the total number of doublons such that $0\leq R_{\bm{k}}\leq 1$ is satisfied, since the number of doublons is not conserved during the dynamics. 
Figure \ref{fig_momdist} shows the dynamics of the doublon momentum distribution for the two initial states as in Fig.~\ref{fig_paircorr}. The build-up dynamics of the distribution at $\bm{k}=\pm\bm{Q}$ clearly signals the formation of $\eta$ pairing.

In Fig.~\ref{fig_momdist}, one can see that the doublons for $\bm{k}\neq\pm\bm{Q}$ are still populated after a long time. The residual doublon distribution is \textit{not} due to imperfection of the $\eta$-pairing formation but arises from a fundamental constraint due to the Fermi statistics. In fact, the doublon momentum distribution can be bounded from above as $\langle d_{\bm{k}}^\dag d_{\bm{k}}\rangle\leq\Lambda_2/2$ \cite{supple}, where $\Lambda_2$ is the maximum eigenvalue of the two-particle reduced density matrix $(\rho_2)_{i\sigma_1 j\sigma_2;k\sigma_3 l\sigma_4}\equiv \langle c_{j\sigma_2}^\dag c_{i\sigma_1}^\dag c_{k\sigma_3}c_{l\sigma_4}\rangle$ \cite{Yang62}. Using  the bound of $\Lambda_2$ from the Pauli exclusion principle \cite{Yang62}, we obtain
\begin{equation}
\langle d_{\bm{k}}^\dag d_{\bm{k}}\rangle\leq\frac{N(2N_s-N+2)}{4N_s},
\label{eq_distbound}
\end{equation}
where the bound is saturated by Yang's $\eta$-pairing state $\ket{\psi_N}$ for $\bm{k}=\bm{Q}$ \cite{supple}. Moreover, provided that the model does not have a dark state other than $\ket{\psi_N}$ and $\ket{\psi_{N_p};\bm{k}_1,\cdots,\bm{k}_{N_{\mathrm{ex}}}}$, we can derive a bound on $R_{\bm{Q}}$ in the steady state as \cite{supple}
\begin{equation}
R_{\bm{Q}}\leq 1-\frac{N}{2N_s}+\frac{1}{N_s}.
\label{eq_Rbound}
\end{equation}
The bound \eqref{eq_Rbound} is consistent with the numerical results in Fig.~\ref{fig_momdist}. If we take the thermodynamic limit $N_s\to\infty$ while keeping the density $n\equiv N/(2N_s)$ constant, the inequality \eqref{eq_Rbound} reduces to $R_{\bm{Q}}\leq 1-n$. Thus, the normalized doublon momentum distribution at momentum $\pm\bm{Q}$ cannot reach unity for a system with nonzero density because of the Pauli exclusion principle, indicating a fundamental distinction between doublons and bosons \cite{supple}.

\textit{Experimental implementation}.--\ 
The validity of the Hubbard model for the description of ultracold fermionic atoms in an optical lattice has been tested by a number of experiments \cite{Esslinger10}. In particular, it has been experimentally verified in Ref.~\cite{Gall20} that a cold-atom quantum simulator of the Hubbard model possesses the Shiba symmetry \cite{Shiba72, Ho09} between the repulsive and attractive Hubbard models, which leads to the $\eta$ SU(2) symmetry when combined with the spin SU(2) symmetry \cite{YangZhang90}.
The Lindblad operator $L_j$ for spontaneous decay from a spin-up state to a spin-down state can be implemented by weakly coupling the spin-up state to an excited state that can decay to the spin-down state [see Fig.~\ref{fig_setup}(b)] \cite{Sandner11, supple}. 
With large detuning of the laser that couples the spin-up state and the excited state, the latter can be adiabatically eliminated \cite{Pichler10, Sarkar14, Reiter12}, and the model \eqref{eq_master} is obtained \cite{supple}.  
To avoid recoil of atoms that induces transitions to neighboring sites or higher bands, a sufficiently deep optical lattice should be used \cite{supple, Pichler10, Sarkar14}. 

In this implementation, a spontaneous decay from the excited state to the spin-up state leads to an additional Lindblad operator $L_j^{(d)}=\sqrt{\gamma_d}n_{j\up}$, which causes dephasing of atoms. Since the inverse of the dephasing rate $\gamma_d$ determines the lifetime of the $\eta$-pairing states as confirmed numerically \cite{supple}, the condition $\gamma_d \ll \gamma,t$ needs to be met.
A possible candidate that achieves this condition is alkaline-earth-like fermionic atoms such as $^{171}$Yb, $^{173}$Yb, and $^{87}$Sr \cite{supple}. We can implement the Fermi-Hubbard model by trapping spin-polarized atoms in the ground $^1S_0$ state and in the metastable $^3P_0$ state in a magic-wavelength optical lattice \cite{Gorshkov10}. We can also realize an effective spontaneous decay by coupling the $^3P_0$ state to a particular hyperfine state in the $^1P_1$ state that can quickly decay to the $^1S_0$ state \cite{Sandner11}. Since the spontaneous decay from the $^1P_1$ state to the $^3P_0$ state is negligible in the experimental time scale, the condition $\gamma_d\ll \gamma,t$ can well be satisfied.


\textit{Summary}.--\ 
In this Letter, we have proposed a nonequilibrium pairing mechanism for fermionic atoms using spontaneous emission of light. 
The nonequilibrium $\eta$ pairing has been shown to be formed through a non-BCS mechanism, where the Pauli blocking of radiative decay of on-site fermion pairs and the destructive interference between doublon-decay processes play vital roles. 
We have shown that $\eta$-pairing correlations in the steady state are supported by high density of doublons in the initial state, which indicates that coherent $\eta$ pairs can be created at temperatures comparable to or even higher than the Fermi temperature, as long as the single-band approximation is valid. 
Since charge-neutral atoms do not have the dynamical instability of $\eta$-pairing states due to coupling with dynamical electromagnetic fields as shown recently \cite{Tsuji21}, the scheme proposed in this Letter for ultracold atoms serves as a promising candidate for observation of the $\eta$-pairing state well above the BCS critical temperature.

\begin{acknowledgments}
This work was supported by KAKENHI (Grant Nos.~JP18H01140, JP18H01145, JP19H01838, JP20K03811, and JP20K14383) and a Grant-in-Aid for Scientific Research on Innovative Areas (KAKENHI Grant No.~JP15H05855) from the Japan Society for the Promotion of Science. 
\end{acknowledgments}

\bibliography{dark_eta_ref.bib}


\clearpage

\renewcommand{\thesection}{S\arabic{section}}
\renewcommand{\theequation}{S\arabic{equation}}
\setcounter{equation}{0}
\renewcommand{\thefigure}{S\arabic{figure}}
\setcounter{figure}{0}

\onecolumngrid
\appendix
\begin{center}
\large{Supplemental Material for}\\
\textbf{``$\eta$ Pairing of Light-Emitting Fermions: Nonequilibrium Pairing Mechanism at High Temperatures''}
\end{center}


\section{Eliminating the energy difference between spin states}

We show that the energy difference $\delta$ between internal spin states in the model \eqref{eq_Hubbard} does not affect the dynamics of the system. We decompose the Hamiltonian as
\begin{align}
H&=H_{0}+H_\delta,\label{eq_Hdecomp}\\
H_0&\equiv-t\sum_{\langle i,j\rangle}\sum_{\sigma}(c_{i\sigma}^\dag c_{j\sigma}+\mathrm{H.c.})+U\sum_j n_{j\up}n_{j\down},\\
H_\delta&\equiv\delta\sum_j(n_{j\up}-n_{j\down}),
\end{align}
and introduce the interaction representation
\begin{equation}
\tilde{\rho}(\tau)\equiv e^{iH_\delta\tau}\rho(\tau)e^{-iH_\delta\tau}.
\label{eq_deltarep}
\end{equation}
Substituting Eqs.~\eqref{eq_Hdecomp}-\eqref{eq_deltarep} into the quantum master equation \eqref{eq_master}, we obtain
\begin{equation}
\frac{d\tilde{\rho}}{d\tau}=-i[H_0,\tilde{\rho}]+\sum_j(L_j\tilde{\rho} L_j^\dag-\frac{1}{2}\{ L_j^\dag L_j,\tilde{\rho}\}).
\end{equation}
For a physical quantity $\mathcal{O}$ that commutes with the magnetization operator $S_z$, we have
\begin{equation}
\mathrm{Tr}[\mathcal{O}\tilde{\rho}(\tau)]=\mathrm{Tr}[\mathcal{O}\rho(\tau)].
\end{equation}
Thus, the expectation value of an observable is independent of $\delta$ and therefore we may set $\delta=0$ without loss of generality.


\section{Destructive interference between doublon-decay processes}
Here we show that the probability amplitude of decay processes of doublons vanishes for the $\eta$-pairing state. We define a hopping operator for spin-$\sigma$ particles by $T_{ij\sigma}\equiv -t(c_{i\sigma}^\dag c_{j\sigma}+\mathrm{H.c.})$, where $i$ and $j$ denote a pair of nearest-neighbor sites. Then, the commutation relations of the hopping operator and the $\eta$ operator are calculated as
\begin{equation}
[T_{ij\up},\eta^+]=-te^{i\bm{Q}\cdot\bm{R}_j}(c_{i\up}^\dag c_{j\down}^\dag + e^{i\bm{Q}\cdot(\bm{R}_i-\bm{R}_j)}c_{j\up}^\dag c_{i\down}^\dag),
\label{eq_etaTup}
\end{equation}
and
\begin{equation}
[T_{ij\down},\eta^+]=-te^{i\bm{Q}\cdot\bm{R}_j}(e^{i\bm{Q}\cdot(\bm{R}_i-\bm{R}_j)}c_{i\up}^\dag c_{j\down}^\dag + c_{j\up}^\dag c_{i\down}^\dag).
\label{eq_etaTdown}
\end{equation}
Since the right-hand sides of Eqs.~\eqref{eq_etaTup} and \eqref{eq_etaTdown} both commute with $\eta^+$, we have
\begin{equation}
[T_{ij\sigma},(\eta^+)^n]=n(\eta^+)^{n-1}[T_{ij\sigma},\eta^+],
\end{equation}
which leads to
\begin{equation}
T_{ij\sigma}\ket{\psi_N}=\frac{N}{2}(\eta^+)^{N/2-1}[T_{ij\sigma},\eta^+]\ket{0},
\label{eq_Tpsi}
\end{equation}
where $\ket{\psi_N}=(\eta^+)^{N/2}\ket{0}$ is Yang's $\eta$-pairing state. 
Here, we invoke the fact that the center-of-mass momentum $\bm{Q}=(\pi,\cdots,\pi)$ leads to $e^{i\bm{Q}\cdot(\bm{R}_i-\bm{R}_j)}=-1$ for any pair of nearest-neighbor sites $i$ and $j$. Thus, from Eqs.~\eqref{eq_etaTup}, \eqref{eq_etaTdown}, and \eqref{eq_Tpsi}, we obtain 
\begin{equation}
T_{ij\up}\ket{\psi_N}=-T_{\ij\down}\ket{\psi_N},
\label{eq_cancel}
\end{equation}
which indicates a vanishing probability amplitude of the doublon-decay processes, i.e., $-t\sum_\sigma(c_{i\sigma}^\dag c_{j\sigma}+\mathrm{H.c.})\ket{\psi_N}=0$. As clearly seen from Eq.~\eqref{eq_cancel}, the physical origin of this vanishing probability amplitude is that the minus sign [$e^{i\bm{Q}\cdot(\bm{R}_i-\bm{R}_j)}=-1$] arising from the $\eta$-pair wave function leads to complete destructive interference between two paths of doublon-decay processes described by $T_{ij\up}$ and $T_{ij\down}$.


\section{Exact Liouvillian eigenmodes}
The Liouvillian superoperator and its conjugate are defined as $\mathcal{L}\rho=-i[H,\rho]+\sum_j(L_j\rho L_j^\dag-\frac{1}{2}\{ L_j^\dag L_j,\rho\})$ and $\mathcal{L}^\dag\rho=i[H,\rho]+\sum_j(L_j^\dag\rho L_j-\frac{1}{2}\{ L_j^\dag L_j,\rho\})$, respectively. 
Suppose that the Liouvillian is diagonalized as $\mathcal{L}\rho_n^{\mathrm{R}}=\lambda_n\rho_n^{\mathrm{R}}$ and $\mathcal{L}^\dag\rho_n^{\mathrm{L}}=\lambda_n^*\rho_n^{\mathrm{L}}$, where $\rho_n^{\mathrm{R}}$ ($\rho_n^{\mathrm{L}}$) denotes the $n$th right (left) eigenmode. Then, the time evolution of a density matrix can be written as
\begin{equation}
\rho(\tau)=\sum_n c_ne^{\lambda_n\tau}\rho_n^{\mathrm{R}},
\end{equation}
where the coefficient reads $c_n=\mathrm{Tr}[(\rho_n^{\mathrm{L}})^\dag \rho(0)]$ if the biorthonormal condition $\mathrm{Tr}[(\rho_m^{\mathrm{L}})^\dag\rho_n^{\mathrm{R}}]=\delta_{m,n}$ is imposed. As mentioned in the main text, the Liouvillian $\mathcal{L}=\mathcal{K}+\mathcal{J}$ can be made to take a triangular-matrix form whose diagonal (off-diagonal) components arise from $\mathcal{K}$ ($\mathcal{J}$). Since eigenvalues of a triangular matrix are given by its diagonal elements, the Liouvillian eigenvalues coincide with those of $\mathcal{K}$. Thus, eigenvalues $\lambda$ of the Liouvillian are given as $\lambda=-i(E_a-E_b^*)$, where $E_a$ and $E_b$ are eigenvalues of the non-Hermitian Hamiltonian
\begin{align}
H_{\mathrm{eff}}=&H-\frac{i}{2}\sum_j L_j^\dag L_j\notag\\
=&-t\sum_{\langle i,j\rangle}\sum_\sigma(c_{i\sigma}^\dag c_{j\sigma}+\mathrm{H.c.})+\left(U+\frac{i\gamma}{2}\right)\sum_j n_{j\up}n_{j\down}
+\left(\delta-\frac{i\gamma}{2}\right)\sum_jn_{j\up}-\delta\sum_jn_{j\down},
\label{eq_Heff}
\end{align}
and the corresponding eigenmodes of the Liouvillian can be obtained from the eigenstates of $H_{\mathrm{eff}}$ and the matrix elements of $L_j$ between them \cite{Torres14, Nakagawa20}.

Single-particle excitations from the $\eta$-pairing state $\ket{\psi_N}$ were constructed by Yang \cite{Yang89}. We define
\begin{gather}
\ket{\zeta_{N,\bm{a}}}\equiv (\eta^+)^{N/2-1}\eta_{\bm{a}}^+\ket{0},\\
\eta_{\bm{a}}^+\equiv\sum_{\bm{k}}e^{-i\bm{k}\cdot \bm{a}}c_{\bm{k}\up}^\dag c_{\bm{Q}-\bm{k}\down}^\dag,
\end{gather}
for $\bm{a}\in\{0,\cdots,L-1\}^d$ with $\bm{a}\neq (0,\cdots,0)$. Here $\ket{\zeta_{N,\bm{a}}}$ is an exact $N$-particle eigenstate of $H_{\mathrm{eff}}$ in any spatial dimension:
\begin{equation}
H_{\mathrm{eff}}\ket{\zeta_{N,\bm{a}}}=\left[\frac{NU}{2}-U-\frac{i\gamma}{2}\right]\ket{\zeta_{N,\bm{a}}},
\end{equation}
which indicates that $\ket{\zeta_{N,\bm{a}}}$ is a single-particle excited state with the same momentum as the $\eta$-pairing state \cite{Yang89}. The gap $\mathrm{Re}[-U-i\gamma/2]=-U$ in the excitation energy is related to the metastability of the $\eta$-pairing state for $U<0$ in an isolated system \cite{Yang89}. 
To promote the eigenstates of the non-Hermitian Hamiltonian to eigenmodes of the Liouvillian, we assume the form of an eigenmode as
\begin{equation}
\rho_{N,\bm{a},\bm{a}'}^{\mathrm{R}}=C_{N,\bm{a},\bm{a}'}\ket{\zeta_{N,\bm{a}}}\bra{\zeta_{N,\bm{a}'}}+\sum_{\bm{k}_1,\bm{k}_2,\bm{k}'_1,\bm{k}'_2}D_{\bm{k}_1,\bm{k}_2,\bm{k}'_1,\bm{k}'_2}^{(N,\bm{a},\bm{a}')}\ket{\psi_{N-2};\bm{k}_1,\bm{k}_2}\bra{\psi_{N-2};\bm{k}'_1,\bm{k}'_2},
\label{eq_single_eigen}
\end{equation}
which is obtained from the general form of solutions in Refs.~\cite{Torres14, Nakagawa20}. 
Note that $L_j\ket{\psi_{N-2};\bm{k}_1,\bm{k}_2}=0$. 
Substituting Eq.~\eqref{eq_single_eigen} into the eigenvalue equation $\mathcal{L}\rho_{N,\bm{a},\bm{a}'}^{\mathrm{R}}=\lambda_{N,\bm{a},\bm{a}'}\rho_{N,\bm{a},\bm{a}'}^{\mathrm{R}}$, we obtain
\begin{gather}
\lambda_{N,\bm{a},\bm{a}'}=-\gamma,
\label{eq_lambda_1exci}
\end{gather}
and
\begin{align}
D_{\bm{k}_1,\bm{k}_2,\bm{k}'_1,\bm{k}'_2}^{(N,\bm{a},\bm{a}')}=\frac{\gamma C_{N,\bm{a},\bm{a}'}}{N_s(\lambda_{N,\bm{a},\bm{a}'}-\lambda_{\bm{k}_1,\bm{k}_2,\bm{k}'_1,\bm{k}'_2})}\delta_{\bm{k}_1+\bm{k}_2,\bm{k}'_1+\bm{k}'_2}e^{-i(\bm{Q}-\bm{k}_2)\cdot\bm{a}}e^{i(\bm{Q}-\bm{k}'_2)\cdot\bm{a}'},
\end{align}
where $\lambda_{\bm{k}_1,\bm{k}_2,\bm{k}'_1,\bm{k}'_2}=2it\sum_{\mu=1}^d(\cos k_{1\mu}+\cos k_{2\mu}-\cos k'_{1\mu}-\cos k'_{2\mu})$. The normalization constant is given by
\begin{equation}
C_{N,\bm{a},\bm{a}'}=(\braket{\zeta_{N,\bm{a}}|\zeta_{N,\bm{a}}}\braket{\zeta_{N,\bm{a}'}|\zeta_{N,\bm{a}'}})^{-1/2}=\frac{(N_s-N/2-1)!}{N_s(N/2-1)!(N_s-2)!}.
\end{equation}
Since the eigenvalue \eqref{eq_lambda_1exci} does not depend on $N$, $\bm{a}$, and $\bm{a}'$, the single-particle excitations decay at a constant rate $\gamma$.

The $\eta$-pairing state breaks the $\eta$ SU(2) symmetry. This implies that the system hosts collective Nambu-Goldstone modes. An exact expression of the dispersion relation of the collective modes can be obtained for a one-dimensional system by the Bethe ansatz method \cite{Nakagawa20}. We consider an $N$-particle eigenstate of the form $\ket{\phi_N}\equiv(\eta^+)^{N/2-1}\ket{\phi_2}$, where $\ket{\phi_2}$ is a two-particle eigenstate. The Bethe equations for a two-particle system are given by \cite{1dHubbard_book}
\begin{align}
e^{ik_1L}=&\frac{\sin k_1-\sin k_2 +2iu}{\sin k_1 -\sin k_2 -2iu},\label{eq_2Bethe1_2}\\
e^{ik_2L}=&\frac{\sin k_2-\sin k_1 +2iu}{\sin k_2 -\sin k_1 -2iu},\label{eq_2Bethe2_2}
\end{align}
where $k_1$ and $k_2$ denote quasimomenta, and $u\equiv(U+i\gamma/2)/(4t)$. The collective excitations are obtained from string solutions in which the quasimomenta take complex values. Without loss of generality, we assume that $k_1=q_1-i\kappa$ and $k_2=q_2+i\kappa\ (q_1,q_2\in\mathbb{R},\kappa>0)$. Then, taking the limit of $L\to\infty$ in Eqs.~\eqref{eq_2Bethe1_2} and \eqref{eq_2Bethe2_2}, we have
\begin{equation}
\sin k_1 -\sin k_2 =2iu,
\end{equation}
the real and imaginary parts of which read
\begin{align}
\cos\frac{P}{2}\sin\frac{q}{2}\cosh\kappa=&-\mathrm{Im}[u],\label{eq_bound1}\\
\cos\frac{P}{2}\cos\frac{q}{2}\sinh\kappa=&-\mathrm{Re}[u],\label{eq_bound2}
\end{align}
where $P\equiv q_1+q_2$ is the momentum eigenvalue, and $q\equiv q_1-q_2$. Using the solution of Eqs.~\eqref{eq_bound1} and \eqref{eq_bound2}, the two-particle eigenstate $\ket{\phi_2}$ is given by $\ket{\phi_2}=\sum_{x_1,x_2}\psi(x_1,x_2)c_{x_1\up}^\dag c_{x_2\down}^\dag\ket{0}$, where
\begin{equation}
\psi(x_1,x_2)=A[\theta(x_2-x_1)e^{iq_1x_1+iq_2x_2}+\theta(x_1-x_2)e^{iq_1x_2+iq_2x_1}]e^{-\kappa|x_1-x_2|}
\end{equation}
is the Bethe wavefunction \cite{1dHubbard_book}, $A$ is the normalization constant, and $\theta(x)$ is the Heaviside unit-step function. 
The energy eigenvalue $E(P)$ of $\ket{\phi_N}$ is obtained by using that of $\ket{\phi_2}$ and the $\eta$ SU(2) symmetry:
\begin{align}
E(P)=&-2t(\cos k_1 +\cos k_2)-\frac{i\gamma}{2}+\left(\frac{N}{2}-1\right)U\notag\\
=&-4t\cos\frac{P}{2}\cos\left(\frac{q}{2}-i\kappa\right)-\frac{i\gamma}{2}+\left(\frac{N}{2}-1\right)U\notag\\
=&\sqrt{16t^2\cos^2\frac{P}{2}+\left(U+\frac{i\gamma}{2}\right)^2}-\frac{i\gamma}{2}+\left(\frac{N}{2}-1\right)U,
\end{align}
where we take the branch of the square root so as to make its imaginary part positive. 
The Liouvillian eigenmode for a collective excitation with momentum $P$ is given by
\begin{align}
\rho_{N}^{\mathrm{R}}(P)=C_N\ket{\phi_N}\bra{\phi_N}+\sum_{\bm{k}_1,\bm{k}_2,\bm{k}'_1,\bm{k}'_2}D_{\bm{k}_1,\bm{k}_2,\bm{k}'_1,\bm{k}'_2}^{(N)}\ket{\psi_{N-2};\bm{k}_1,\bm{k}_2}\bra{\psi_{N-2};\bm{k}'_1,\bm{k}'_2},
\end{align}
and its eigenvalue is
\begin{align}
\lambda(P)=-i(E(P)-E^*(P))=2\mathrm{Im}\sqrt{16t^2\cos^2\frac{P}{2}+\left(U+\frac{i\gamma}{2}\right)^2}-\gamma.
\end{align}
The coefficients $C_N$ and $D_{\bm{k}_1,\bm{k}_2,\bm{k}'_1,\bm{k}'_2}^{(N)}$ can be determined from the general procedure in Refs.~\cite{Torres14, Nakagawa20}. In particular, for $P\simeq \pi$, the Liouvillian eigenvalue reads
\begin{equation}
\lambda(P)\simeq 2\mathrm{Im}[J](1+\cos P),
\end{equation}
where $J=4t^2/(U+i\gamma/2)$ is an effective exchange coupling between the $\eta$ SU(2) pseudospins. The dispersion relation is gapless and quadratic near $P=\pi$. 
The collective excitations with the quadratic dispersion are understood as pseudospin waves of the $\eta$ SU(2) pseudospins, which are similar to magnons of the ferromagnetic Heisenberg model. 
In fact, by the Shiba transformation, $\eta$ SU(2) pseudospins are mapped to ordinary SU(2) spins and the collective mode corresponds to a ferromagnetic spin wave in a dissipative Hubbard model subject to two-body loss \cite{Nakagawa20}.


\section{Derivation of the bound on the doublon momentum distribution}

Here, we derive the bounds [Eqs.~\eqref{eq_distbound} and \eqref{eq_Rbound} in the main text] on the momentum distribution of doublons. We define a two-particle reduced density matrix $\rho_2$ by
\begin{equation}
(\rho_2)_{i\sigma_1 j\sigma_2,k\sigma_3 l\sigma_4}\equiv\mathrm{Tr}[c_{j\sigma_2}^\dag c_{i\sigma_1}^\dag c_{k\sigma_3}c_{l\sigma_4}\rho]
\end{equation}
for a density matrix $\rho$ of an $N$-particle system \cite{Yang62}. The two-particle reduced density matrix $\rho_2$ is Hermitian and positive semidefinite \cite{Yang62}. Let $\Lambda_2$ and $v=(v_{i\sigma_1j\sigma_2})$ be the maximum eigenvalue and the corresponding eigenvector of $\rho_2$ that satisfy
\begin{equation}
\sum_{k,l}\sum_{\sigma_3,\sigma_4}(\rho_2)_{i\sigma_1 j\sigma_2,k\sigma_3 l\sigma_4}v_{k\sigma_3 l\sigma_4}=\Lambda_2 v_{i\sigma_1j\sigma_2}.
\end{equation}
Then, the inequality
\begin{equation}
\langle f,\rho_2 f\rangle\equiv\sum_{i,j,k,l}\sum_{\sigma_1,\cdots,\sigma_4}f_{i\sigma_1j\sigma_2}^*(\rho_2)_{i\sigma_1 j\sigma_2,k\sigma_3 l\sigma_4}f_{k\sigma_3 l\sigma_4}\leq \Lambda_2
\label{eq_rho2_ineq1}
\end{equation}
holds for any vector $f=(f_{i\sigma_1j\sigma_2})$ which is normalized as $\sum_{i,j}\sum_{\sigma_1,\sigma_2}|f_{i\sigma_1j\sigma_2}|^2=1$. To derive the bound on $\langle d_{\bm{k}}^\dag d_{\bm{k}}\rangle$, we set
\begin{align}
f_{i\sigma_1j\sigma_2}=\frac{1}{\sqrt{2N_s}}\delta_{i,j}e^{-i\bm{k}\cdot\bm{R}_i}(\delta_{\sigma_1,\down}\delta_{\sigma_2,\up}-\delta_{\sigma_1,\up}\delta_{\sigma_2,\down}),
\end{align}
and define
\begin{align}
F\equiv\sum_{i,j}\sum_{\sigma_1,\sigma_2}f_{i\sigma_1j\sigma_2}c_{i\sigma_1}c_{j\sigma_2}
=\frac{\sqrt{2}}{\sqrt{N_s}}\sum_je^{-i\bm{k}\cdot\bm{R}_j}c_{j\down}c_{j\up}.
\end{align}
Then, we have
\begin{align}
\langle d_{\bm{k}}^\dag d_{\bm{k}}\rangle=\frac{1}{N_s}\sum_{i,j}e^{i\bm{k}\cdot(\bm{R}_i-\bm{R}_j)}\mathrm{Tr}[c_{i\up}^\dag c_{i\down}^\dag c_{j\down} c_{j\up}\rho]
=\frac{1}{2}\mathrm{Tr}[F^\dag F\rho]
=\frac{1}{2}\langle f,\rho_2 f\rangle
\leq \frac{\Lambda_2}{2}.
\end{align}
The bound on $\Lambda_2$ was derived by Yang \cite{Yang62}:
\begin{equation}
\Lambda_2\leq\frac{N(M-N+2)}{M},
\label{eq_Yang_bound}
\end{equation}
where $M$ denotes the number of single-particle states of fermions. For our case, $M=2N_s$, and thus we obtain the bound on the number of doublons at momentum $\bm{k}$:
\begin{equation}
\langle d_{\bm{k}}^\dag d_{\bm{k}}\rangle\leq \frac{N(2N_s-N+2)}{4N_s}=\frac{N}{2}-\frac{N(N-2)}{4N_s},
\label{eq_distbound_supple}
\end{equation}
which indicates that an $N$-particle system of $N/2(\geq 2)$ doublons cannot condense all doublons into a single momentum state. This bound \eqref{eq_distbound_supple} holds for an arbitrary density matrix $\rho$. The physical origin of this bound can be understood from the Pauli exclusion principle; fermions cannot be condensed without forming pairs and therefore the maximum eigenvalue of the one-particle (two-particle) reduced density matrix cannot be of the order of $N$ ($N^2$) \cite{Yang62}.

For Yang's $\eta$-pairing state $\ket{\psi_N}=(\eta^+)^{N/2}\ket{0}$, the pair correlation function can be calculated as $\langle c_{i\up}^\dag c_{i\down}^\dag c_{j\down} c_{j\up}\rangle=\frac{(N/2)(N_s-N/2)}{N_s(N_s-1)}e^{i\bm{Q}\cdot(\bm{R}_i-\bm{R}_j)}$ \cite{Yang89}. Thus, we have
\begin{align}
\langle d_{\bm{Q}}^\dag d_{\bm{Q}}\rangle=&\frac{1}{N_s}\sum_j\langle n_{j\up}n_{j\down}\rangle+\frac{1}{N_s}\sum_{i\neq j}\langle c_{i\up}^\dag c_{i\down}^\dag c_{j\down} c_{j\up}\rangle e^{i\bm{Q}\cdot(\bm{R}_i-\bm{R}_j)}\notag\\
=&\frac{N/2}{N_s}+\frac{1}{N_s}\frac{(N/2)(N_s-N/2)}{N_s(N_s-1)}\times N_s(N_s-1)\notag\\
=&\frac{N(2N_s-N+2)}{4N_s}.
\label{eq_eta_dQ}
\end{align}
Thus, Yang's $\eta$-pairing state $\ket{\psi_N}$ achieves the bound \eqref{eq_distbound_supple} for $\bm{k}=\bm{Q}$, and describes the maximal condensation of doublons at momentum $\bm{Q}$. The effect of the Pauli exclusion principle in the momentum distribution can clearly be seen by comparing the commutation relations of annihilation and creation operators of doublons with those of bosons; if $b_{\bm{k}}$ is an annihilation operator of a boson that satisfies the canonical commutation relation $[b_{\bm{k}},b_{\bm{k}'}^\dag]=\delta_{\bm{k},\bm{k}'}$, the momentum distribution of bosons for a state $(b_{\bm{Q}}^\dag)^N\ket{0}$ satisfies $\bra{0}(b_{\bm{Q}})^N b_{\bm{k}}^\dag b_{\bm{k}}(b_{\bm{Q}}^\dag)^N\ket{0}=\bra{0}(b_{\bm{Q}})^N b_{\bm{k}}^\dag (b_{\bm{Q}}^\dag)^N b_{\bm{k}}\ket{0}=0$ for $\bm{k}\neq\bm{Q}$, and thus all bosons are condensed at momentum $\bm{Q}$. However, for the annihilation operator of a doublon, the commutation relations read
\begin{gather}
[d_{\bm{k}},d_{\bm{k}'}]=[d_{\bm{k}}^\dag,d_{\bm{k}'}^\dag]=0,\\
[d_{\bm{k}},d_{\bm{k}'}^\dag]=\delta_{\bm{k},\bm{k}'}-\frac{1}{N_s}\sum_j(n_{j\up}+n_{j\down})e^{-i(\bm{k}-\bm{k}')\cdot\bm{R}_j},
\end{gather}
and do not satisfy the canonical commutation relations for bosons because of the Pauli exclusion principle. For this reason, doublons can be distributed at momentum $\bm{k}\neq\bm{Q}$ in the $\eta$-pairing state $\ket{\psi_N}=(d_{\bm{Q}}^\dag)^{N/2}\ket{0}$.

The momentum distribution of doublons for the $\eta$-pairing state $\ket{\psi_{2N_p};\bm{k}_1,\cdots,\bm{k}_{N_{\mathrm{ex}}}}=(\eta^+)^{N_p}c_{\bm{k}_1\down}^\dag \cdots c_{\bm{k}_{N_{\mathrm{ex}}}\down}^\dag\ket{0}$ with excess spin-down particles can also be calculated. We note that
\begin{align}
\bm{\eta}^2\ket{\psi_{2N_p};\bm{k}_1,\cdots,\bm{k}_{N_{\mathrm{ex}}}}=&\frac{N_s-N_{\mathrm{ex}}}{2}\left(\frac{N_s-N_{\mathrm{ex}}}{2}+1\right)\ket{\psi_{2N_p};\bm{k}_1,\cdots,\bm{k}_{N_{\mathrm{ex}}}},\\
\eta^z\ket{\psi_{2N_p};\bm{k}_1,\cdots,\bm{k}_{N_{\mathrm{ex}}}}=&\frac{2N_p+N_{\mathrm{ex}}-N_s}{2}\ket{\psi_{2N_p};\bm{k}_1,\cdots,\bm{k}_{N_{\mathrm{ex}}}},
\end{align}
where $\bm{\eta}^2=(\eta^x)^2+(\eta^y)^2+(\eta^z)^2$. Thus, using $\eta^+\eta^-=\bm{\eta}^2-(\eta^z)^2+\eta^z$, we obtain
\begin{align}
\langle d_{\bm{Q}}^\dag d_{\bm{Q}}\rangle=&\frac{1}{N_s}\frac{\bra{\psi_{2N_p};\bm{k}_1,\cdots,\bm{k}_{N_{\mathrm{ex}}}}\eta^+\eta^-\ket{\psi_{2N_p};\bm{k}_1,\cdots,\bm{k}_{N_{\mathrm{ex}}}}}{\braket{\psi_{2N_p};\bm{k}_1,\cdots,\bm{k}_{N_{\mathrm{ex}}}|\psi_{2N_p};\bm{k}_1,\cdots,\bm{k}_{N_{\mathrm{ex}}}}}\notag\\
=&\frac{1}{N_s}\frac{N_s-N_{\mathrm{ex}}}{2}\left(\frac{N_s-N_{\mathrm{ex}}}{2}+1\right)-\frac{1}{N_s}\frac{2N_p+N_{\mathrm{ex}}-N_s}{2}\left(\frac{2N_p+N_{\mathrm{ex}}-N_s}{2}-1\right)\notag\\
=&\frac{N_p(N_s-N_p-N_{\mathrm{ex}}+1)}{N_s},
\label{eq_etaex_dQ}
\end{align}
which is lower than that for $\ket{\psi_{2N_p}}$ due to the Pauli exclusion between doublons and excess spin-down particles.

In the main text, we have calculated the doublon momentum distribution $R_{\bm{k}}=\langle d_{\bm{k}}^\dag d_{\bm{k}}\rangle/\sum_{\bm{k}}\langle d_{\bm{k}}^\dag d_{\bm{k}}\rangle$ that is normalized by the total number of doublons. The bound of the normalized momentum distribution can be obtained under the assumption that a steady state $\rho_{\mathrm{SS}}$ is a statistical mixture of the $\eta$-pairing states $\ket{\psi_N}$ and $\ket{\psi_{2N_p};\bm{k}_1,\cdots,\bm{k}_{N_{\mathrm{ex}}}}$ as
\begin{equation}
\rho_{\mathrm{SS}}=\sum_{N_{\mathrm{ex}}=0}^N p(N_{\mathrm{ex}})\rho\left(\frac{N-N_{\mathrm{ex}}}{2},N_{\mathrm{ex}}\right),
\end{equation}
where
\begin{gather}
\rho(N/2,0)\equiv\frac{\ket{\psi_N}\bra{\psi_N}}{\braket{\psi_N|\psi_N}},\\
\rho(N_p,N_{\mathrm{ex}})\equiv\sum_{\bm{k}_1,\cdots,\bm{k}_{N_{\mathrm{ex}}}}\frac{p(N_p,\bm{k}_1,\cdots,\bm{k}_{N_{\mathrm{ex}}})}{\sum_{\bm{k}_1,\cdots,\bm{k}_{N_{\mathrm{ex}}}}p(N_p,\bm{k}_1,\cdots,\bm{k}_{N_{\mathrm{ex}}})}\frac{\ket{\psi_{2N_p};\bm{k}_1,\cdots,\bm{k}_{N_{\mathrm{ex}}}}\bra{\psi_{2N_p};\bm{k}_1,\cdots,\bm{k}_{N_{\mathrm{ex}}}}}{\braket{\psi_{2N_p};\bm{k}_1,\cdots,\bm{k}_{N_{\mathrm{ex}}}|\psi_{2N_p};\bm{k}_1,\cdots,\bm{k}_{N_{\mathrm{ex}}}}}\ \ (\mathrm{for}\ N_{\mathrm{ex}}\geq 1),
\end{gather}
and the probability distribution satisfies $p(N_p,\bm{k}_1,\cdots,\bm{k}_{N_{\mathrm{ex}}})\geq 0, p(N_{\mathrm{ex}})\geq 0$, and $\sum_{N_{\mathrm{ex}}=0}^N p(N_{\mathrm{ex}})=1$. Note that the dynamics under consideration conserves the total number of particles, and $N_{\mathrm{ex}}$ should be even if the total number of particles is even. Then, from Eqs.~\eqref{eq_eta_dQ} and \eqref{eq_etaex_dQ}, the doublon momentum distribution for the steady state is given by
\begin{align}
\langle d_{\bm{Q}}^\dag d_{\bm{Q}}\rangle=&\mathrm{Tr}[d_{\bm{Q}}^\dag d_{\bm{Q}}\rho_{\mathrm{SS}}]\notag\\
=&\sum_{N_{\mathrm{ex}}=0}^N p(N_{\mathrm{ex}})\frac{N-N_{\mathrm{ex}}}{2N_s}\left(N_s-\frac{N}{2}-\frac{N_{\mathrm{ex}}}{2}+1\right)\notag\\
=&\frac{N_s-N/2+1}{N_s}\sum_{N_{\mathrm{ex}}=0}^N p(N_{\mathrm{ex}})\frac{N-N_{\mathrm{ex}}}{2}-\frac{1}{N_s}\sum_{N_{\mathrm{ex}}=0}^N p(N_{\mathrm{ex}})\frac{N-N_{\mathrm{ex}}}{2}\frac{N_{\mathrm{ex}}}{2}.
\end{align}
The total number of doublons in the steady state is
\begin{align}
\sum_{\bm{k}}\langle d_{\bm{k}}^\dag d_{\bm{k}}\rangle=\sum_j\langle n_{j\up}n_{j\down}\rangle=\sum_j\mathrm{Tr}[n_{j\up}n_{j\down}\rho_{\mathrm{SS}}]
=\sum_{N_{\mathrm{ex}}=0}^N p(N_{\mathrm{ex}})\frac{N-N_{\mathrm{ex}}}{2}.
\end{align}
Thus, we obtain
\begin{align}
R_{\bm{Q}}=&\frac{\langle d_{\bm{Q}}^\dag d_{\bm{Q}}\rangle}{\sum_{\bm{k}}\langle d_{\bm{k}}^\dag d_{\bm{k}}\rangle}\notag\\
=&\frac{N_s-N/2+1}{N_s}-\frac{(1/N_s)\sum_{N_{\mathrm{ex}}=0}^N p(N_{\mathrm{ex}})(N-N_{\mathrm{ex}})/2\cdot N_{\mathrm{ex}}/2}{\sum_{N_{\mathrm{ex}}=0}^N p(N_{\mathrm{ex}})(N-N_{\mathrm{ex}})/2}\notag\\
\leq& \frac{N_s-N/2+1}{N_s}\notag\\
=&1-\frac{N}{2N_s}+\frac{1}{N_s},
\label{eq_Rupper}
\end{align}
which completes the proof of the bound \eqref{eq_Rbound} in the main text.


\section{Details of the experimental implementation}
Here, we describe an experimental situation that leads to the quantum master equation \eqref{eq_master} in the main text. We consider fermionic atoms with a lambda-type level structure shown in Fig.~\ref{fig_setup}(b) in the main text. The two low-lying energy levels are denoted by $\ket{g,\sigma}\ (\sigma=\up,\down)$, and the highest energy level is denoted by $\ket{e}$. We assume that the $\ket{g,\up}$ state is coupled to the $\ket{e}$ state by a coherent laser field, which is described by a classical electromagnetic field, and that the $\ket{e}$ state can decay to the $\ket{g,\sigma}$ state via spontaneous emission of a photon. Then, the Hamiltonian $H_{\mathrm{tot}}$ of the total system is given by
\begin{align}
H_{\mathrm{tot}}=&H_{\mathrm{A}}+H_{\mathrm{I}}+H_{\mathrm{ph}},\\
H_{\mathrm{A}}=&\int d^3\bm{x}\sum_{\sigma=\up,\down}\psi_\sigma^\dag(\bm{x})\left(\frac{-\nabla^2}{2m}+V(\bm{x})\right)\psi_\sigma(\bm{x})+g_0\int d^3\bm{x}\psi_\up^\dag(\bm{x})\psi_\down^\dag(\bm{x}) \psi_\down(\bm{x})\psi_\up(\bm{x})\notag\\
&+\delta_0\int d^3\bm{x}(\psi_\up^\dag(\bm{x})\psi_\up(\bm{x})-\psi_\down^\dag(\bm{x})\psi_\down(\bm{x}))
+(\Delta+\delta_0)\int d^3\bm{x}\psi_e^\dag(\bm{x})\psi_e(\bm{x})
+\int d^3\bm{x}(\Omega(\bm{x})\psi_e^\dag(\bm{x})\psi_\up(\bm{x})+\mathrm{H.c.}),\\
H_{\mathrm{I}}=&-\sum_{\sigma=\up,\down}\int d^3\bm{x}(\bm{d}_{\sigma}\cdot\bm{E}^{(+)}(\bm{x})\psi_e^\dag(\bm{x})\psi_\sigma(\bm{x})+\mathrm{H.c.}),\\
H_{\mathrm{ph}}=&\sum_{\bm{k},\lambda}\omega_{\bm{k}}a_{\bm{k},\lambda}^\dag a_{\bm{k},\lambda},
\end{align}
where $\psi_\sigma(\bm{x})$ ($\psi_e(\bm{x})$) is the field operator of atoms in the $\ket{g,\sigma}$ ($\ket{e}$) state, $m$ is the mass of an atom, $V(\bm{x})$ is an optical lattice potential, $2\delta_0$ is the energy difference between the $\ket{g,\up}$ and $\ket{g,\down}$ states, $g_0$ is the contact interaction strength, $\Delta$ is the detuning of a laser that couples the $\ket{g,\up}$ state to the $\ket{e}$ state with the Rabi frequency $\Omega(\bm{x})$, $\bm{d}_\sigma$ is the electric-dipole matrix element between the states $\ket{g,\sigma}$ and $\ket{e}$, $a_{\bm{k},\lambda}$ is the annihilation operator of a photon field with momentum $\bm{k}$, polarization $\lambda$ and frequency $\omega_{\bm{k}}$, and
\begin{equation}
\bm{E}^{(+)}(\bm{x})=\sum_{\bm{k},\lambda}\varepsilon_{\bm{k}}\bm{e}_{\bm{k},\lambda}e^{i\bm{k}\cdot\bm{x}}a_{\bm{k},\lambda}
\end{equation}
is the positive-frequency part of the electric-field operator with the polarization vector $\bm{e}_{\bm{k},\lambda}$. Here, the rotating-wave approximation has been applied to the light-matter couplings. The kinetic energy of the $\ket{e}$ state can be neglected if the decay of this state occurs sufficiently faster than the time scale of the atomic gas. 
We assume that the relevant optical frequency for the transition between the $\ket{g,\sigma}$ and $\ket{e}$ states is much higher than the energy scale of the atomic gas, and that the quantized radiation modes are in the vacuum state. Then, we can trace out the photon degrees of freedom using the Born-Markov approximation \cite{BreuerPetruccione}, which leads to the quantum master equation for the density matrix $\rho_{\mathrm{A}}$ of the atomic system as \cite{Lehmberg70, Pichler10, Sarkar14}
\begin{align}
\frac{d\rho_{\mathrm{A}}}{d\tau}=&-i[H_A+H_{\mathrm{dip}},\rho_{\mathrm{A}}]+\mathcal{D}\rho_{\mathrm{A}},
\end{align}
where
\begin{align}
H_{\mathrm{dip}}=&\sum_{\sigma=\up,\down}\frac{\Gamma_\sigma}{2}\int d^3\bm{x}\int d^3\bm{y}G_\sigma(k_\sigma(\bm{x}-\bm{y}))\psi_e^\dag(\bm{x})\psi_\sigma(\bm{x})\psi_\sigma^\dag(\bm{y})\psi_e(\bm{y})
\end{align}
is an effective dipole-dipole interaction,
\begin{align}
\mathcal{D}\rho_{\mathrm{A}}=&\sum_{\sigma=\up,\down}\frac{\Gamma_\sigma}{2}\int d^3\bm{x}\int d^3\bm{y}F_\sigma(k_\sigma(\bm{x}-\bm{y}))\Bigl(2\psi_\sigma^\dag(\bm{x})\psi_e(\bm{x})\rho_{\mathrm{A}}\psi_e^\dag(\bm{y})\psi_\sigma(\bm{y})\notag\\
&-\psi_e^\dag(\bm{x})\psi_\sigma(\bm{x})\psi_\sigma^\dag(\bm{y})\psi_e(\bm{y})\rho_{\mathrm{A}}-\rho_{\mathrm{A}}\psi_e^\dag(\bm{x})\psi_\sigma(\bm{x})\psi_\sigma^\dag(\bm{y})\psi_e(\bm{y})\Bigr)
\end{align}
describes spontaneous-emission processes, 
$\Gamma_\sigma$ is the rate of spontaneous decay from the $\ket{e}$ state to the $\ket{g,\sigma}$ state, and $\omega_\sigma=ck_\sigma$ is the energy difference between those states. 
Here, the functions $G(\bm{\xi})$ and $F(\bm{\xi})$ are given from a correlation function of the vacuum radiation modes
\begin{align}
\int_0^\infty ds\langle(\bm{d}_\sigma\cdot\bm{E}^{(+)}(\bm{x},s))(\bm{d}_\sigma\cdot\bm{E}^{(+)}(\bm{y},0))^\dag\rangle e^{i\omega_\sigma s}=\frac{\Gamma_\sigma}{2}\Bigl(F_\sigma(k_\sigma(\bm{x}-\bm{y}))+iG_\sigma(k_\sigma(\bm{x}-\bm{y}))\Bigr)
\end{align}
as \cite{Lehmberg70, Pichler10, Sarkar14}
\begin{align}
F_\sigma(\bm{\xi})=&\frac{3}{8\pi}\int d\Omega_k (1-|\hat{\bm{k}}\cdot\hat{\bm{d}}_\sigma|^2)e^{-i\hat{\bm{k}}\cdot\bm{\xi}},\\
G_\sigma(\bm{\xi})=&-\frac{1}{\xi^3}\mathcal{P}\int\frac{d\zeta}{2\pi}\frac{\zeta^3}{\zeta-\xi}F_\sigma(\zeta\bm{\xi}/\xi),&
\end{align}
where $\hat{\bm{k}}\equiv \bm{k}/|\bm{k}|,\hat{\bm{d}}_\sigma\equiv \bm{d}_\sigma/|\bm{d}_\sigma|$, $\int d\Omega_k$ is the integral over the direction of $\hat{\bm{k}}$, and $\mathcal{P}$ denotes the principal value integral. In deriving the above quantum master equation, we have assumed that photons emitted in the $\ket{e}\to\ket{g,\up}$ process and those emitted in the $\ket{e}\to\ket{g,\down}$ process can be distinguished from their frequencies or polarizations.

If the detuning $\Delta$ is much larger than the other energy scales of the atomic gas, i.e., $|\Delta|\gg\Gamma_\sigma, E_R,|g_0|,|\delta_0|,|\Omega(\bm{x})|$ ($E_R$ is the recoil energy), the excited state $\ket{e}$ can be adiabatically eliminated \cite{Pichler10, Sarkar14}. Using the second-order perturbation theory \cite{Reiter12}, we obtain
\begin{align}
\frac{d\rho}{d\tau}=&-i[H_g+H_{g,\mathrm{dip}},\rho]+\sum_{\sigma=\up,\down}\int d^3\bm{x}\int d^3\bm{y}\frac{\Gamma_\sigma\Omega(\bm{x})\Omega^*(\bm{y})}{2\Delta^2}F_\sigma(k_\sigma(\bm{x}-\bm{y}))\Bigl(2\psi_\sigma^\dag(\bm{x})\psi_\up(\bm{x})\rho\psi_\up^\dag(\bm{y})\psi_\sigma(\bm{y})\notag\\
&-\psi_\up^\dag(\bm{x})\psi_\sigma(\bm{x})\psi_\sigma^\dag(\bm{y})\psi_\up(\bm{y})\rho-\rho\psi_\up^\dag(\bm{x})\psi_\sigma(\bm{x})\psi_\sigma^\dag(\bm{y})\psi_\up(\bm{y})\Bigr),
\label{eq_master_elim}
\end{align}
where $\rho\equiv P_g\rho_{\mathrm{A}}P_g$ is the density matrix of the atomic gas with $P_g$ being a projection operator onto the subspace that does not contain atoms in the $\ket{e}$ state, 
\begin{align}
H_g=&\int d^3\bm{x}\sum_{\sigma=\up,\down}\psi_\sigma^\dag(\bm{x})\left(\frac{-\nabla^2}{2m}+V(\bm{x})\right)\psi_\sigma(\bm{x})+g_0\int d^3\bm{x}\psi_\up^\dag(\bm{x})\psi_\down^\dag(\bm{x}) \psi_\down(\bm{x})\psi_\up(\bm{x})\notag\\
&+\delta_0\int d^3\bm{x}(\psi_\up^\dag(\bm{x})\psi_\up(\bm{x})-\psi_\down^\dag(\bm{x})\psi_\down(\bm{x}))-\int d^3\bm{x}\frac{|\Omega(\bm{x})|^2}{\Delta}\psi_\up^\dag(\bm{x})\psi_\up(\bm{x}),
\end{align}
and
\begin{align}
H_{g,\mathrm{dip}}=\sum_{\sigma=\up,\down}\int d^3\bm{x}\int d^3\bm{y}\frac{\Gamma_\sigma\Omega(\bm{x})\Omega^*(\bm{y})}{2\Delta^2}G_\sigma(k_\sigma(\bm{x}-\bm{y}))\psi_\up^\dag(\bm{x})\psi_\sigma(\bm{x})\psi_\sigma^\dag(\bm{y})\psi_\up(\bm{y}).
\end{align}
Then, substituting the expansion $\psi_\sigma(\bm{x})=\sum_{j}\sum_{n}w_n(\bm{x}-\bm{x}_j)c_{j\sigma n}$ with the Wannier function $w_n(\bm{x})$ of the $n$th band of the optical lattice into Eq.~\eqref{eq_master_elim}, we obtain
\begin{align}
\frac{d\rho}{d\tau}=&-i[H_{\mathrm{multi}},\rho]+\sum_{\sigma=\up,\down}\sum_{j_1,\cdots,j_4}\sum_{n_1,\cdots,n_4}\frac{\gamma_{\sigma;j_1,j_2,j_3,j_4}^{(n_1,n_2,n_3,n_4)}}{2}(2c_{j_1\sigma n_1}^\dag c_{j_2\up n_2}\rho c_{j_3\up n_3}^\dag c_{j_4\sigma n_4}\notag\\
&-c_{j_1\up n_1}^\dag c_{j_2\sigma n_2}c_{j_3\sigma n_3}^\dag c_{j_4\up n_4}\rho-\rho c_{j_1\up n_1}^\dag c_{j_2\sigma n_2}c_{j_3\sigma n_3}^\dag c_{j_4\up n_4}),
\end{align}
where
\begin{align}
H_{\mathrm{multi}}=&-\sum_{i,j}\sum_{\sigma=\up,\down}\sum_n t_{ij}^{(n)}c_{i\sigma n}^\dag c_{j\sigma n}
+\sum_{j_1,j_2,j_3,j_4}\sum_{\sigma}\sum_{n_1,n_2,n_3,n_4}U_{\sigma;j_1,j_2,j_3,j_4}^{(n_1,n_2,n_3,n_4)}c_{j_1\up n_1}^\dag c_{j_2\sigma n_2}^\dag c_{j_3\sigma n_3}c_{j_4\up n_4}\notag\\
&+\sum_j\sum_\sigma\sum_n\delta^{(n)}_\sigma c_{j\sigma n}^\dag c_{j\sigma n}
\end{align}
is the multiband tight-binding Hamiltonian, and
\begin{align}
\gamma_{\sigma;j_1,j_2,j_3,j_4}^{(n_1,n_2,n_3,n_4)}=\int d^3\bm{x}\int d^3\bm{y}\frac{\Gamma_\sigma\Omega(\bm{x})\Omega^*(\bm{y})}{\Delta^2}F_\sigma(k_\sigma(\bm{x}-\bm{y}))w_{n_1}^*(\bm{x}-\bm{x}_{j_1})w_{n_2}(\bm{x}-\bm{x}_{j_2})w^*_{n_3}(\bm{y}-\bm{x}_{j_3})w_{n_4}(\bm{y}-\bm{x}_{j_4}).
\label{eq_multigamma}
\end{align}
When the optical lattice potential is sufficiently deep, the nearest-neighbor hopping and the on-site interaction are dominant and the Hamiltonian $H_{\mathrm{multi}}$ is well approximated by the single-band Hubbard model $H$ [Eq.~\eqref{eq_Hubbard} in the main text] \cite{Esslinger10}. Similarly, the spontaneous-emission processes can be neglected except for on-site ones and we may set $j_1=j_2=j_3=j_4$ in Eq.~\eqref{eq_multigamma}. 
In addition, it is shown from Eq.~\eqref{eq_multigamma} that the inter-band processes are suppressed by a factor of $\eta_\up^2+\eta_\sigma^2$, where $\eta_\sigma\equiv k_\sigma a$ is the Lamb-Dicke parameter with $a$ being the localization length of the Wannier function \cite{Pichler10, Sarkar14}. Hence, for a sufficiently deep optical lattice with small $\eta_\up$ and $\eta_\down$, we obtain
\begin{align}
\frac{d\rho}{d\tau}=&-i[H,\rho]+\sum_j(L_j\rho L_j^\dag-\frac{1}{2}\{ L_j^\dag L_j,\rho\})
+\sum_j(L_j^{(d)}\rho (L_j^{(d)})^\dag-\frac{1}{2}\{ (L_j^{(d)})^\dag L_j^{(d)},\rho\}),
\label{eq_master_deph}
\end{align}
where $L_j=\sqrt{\gamma}c_{j\down}^\dag c_{j\up},\gamma=\gamma_{\down;j,j,j,j}^{(0,0,0,0)},L_j^{(d)}=\sqrt{\gamma_d}n_{j\up}, \gamma_d=\gamma_{\up;j,j,j,j}^{(0,0,0,0)}$, and the lowest-band index is denoted by $n=0$. The Lindblad operator $L_j$ describes an effective decay process from the $\ket{g,\up}$ state to the $\ket{g,\down}$ state, and $L_j^{(d)}$ leads to dephasing of spin-up atoms due to photon scattering. Note that the ratio between the effective decay rate $\gamma$ and the tunneling amplitude $t$ reads
\begin{equation}
\frac{\gamma}{t}=\frac{\Gamma_\down}{t}\frac{\Omega^2}{\Delta^2},
\end{equation}
where $\Omega^2\equiv \int d^3\bm{x}\int d^3\bm{y}^3\Omega(\bm{x})\Omega^*(\bm{y})F(k_\down(\bm{x}-\bm{y}))|w_0(\bm{x}-\bm{x}_j)|^2|w_0(\bm{y}-\bm{x}_j)|^2$, and this ratio is not necessarily small if $\Gamma_\down/t$ is large. Thus, if the condition $\gamma_d\ll\gamma,t$ is satisfied, the model in the main text and hence the $\eta$-pairing states can be realized.

The condition $\gamma_d\ll \gamma$ requires $\Gamma_\up\ll\Gamma_\down$. To achieve this, we propose to use alkaline-earth(-like) fermionic atoms such as $^{171}$Yb, $^{173}$Yb, and $^{87}$Sr \cite{Sandner11}. These atoms have the electronic ground state  $^1S_0$ and a metastable excited state $^3P_0$. First, ground-state atoms in two particular hyperfine states $\ket{^1S_0,\sigma}\ (\sigma=\up,\down)$ are prepared in some initial state in an optical lattice. Subsequently, atoms in the $\ket{^1S_0,\down}$ state are excited to the $\ket{^3P_0,\up}$ state by using a $\pi$-pulse laser. When an optical lattice is created by a laser at a magic wavelength, the lattice potentials for the $^1S_0$ and $^3P_0$ states are the same and thus those atoms are described by the Hubbard model with the following identifications: $\ket{g,\up}=\ket{^1S_0,\up}$ and $\ket{g,\down}=\ket{^3P_0,\up}$ \cite{Gorshkov10}. Since the lifetime of the $^3P_0$ state is sufficiently longer than the typical experimental time scale, spontaneous decay from the $^3P_0$ state to the $^1S_0$ state is negligible. Then, the effective decay process described above is induced by coupling the $\ket{^3P_0,\up}$ state with the $\ket{e}=\ket{^1P_1,\up}$ state that can quickly decay to the $\ket{^1S_0,\up}$ state \cite{Sandner11}. The nuclear-spin-changing decay processes can be suppressed at a high magnetic field \cite{Sandner11}. Since the magnetic-dipole-allowed transition from the $\ket{^3P_0,\up}$ state to the $\ket{^1P_1,\up}$ state is forbidden for the electric dipole coupling, the spontaneous decay between these states is negligible, and thus $\Gamma_\up\ll\Gamma_\down$ is satisfied.


\section{Dependence of pair correlations on model parameters}
\begin{figure}
\includegraphics[width=14cm]{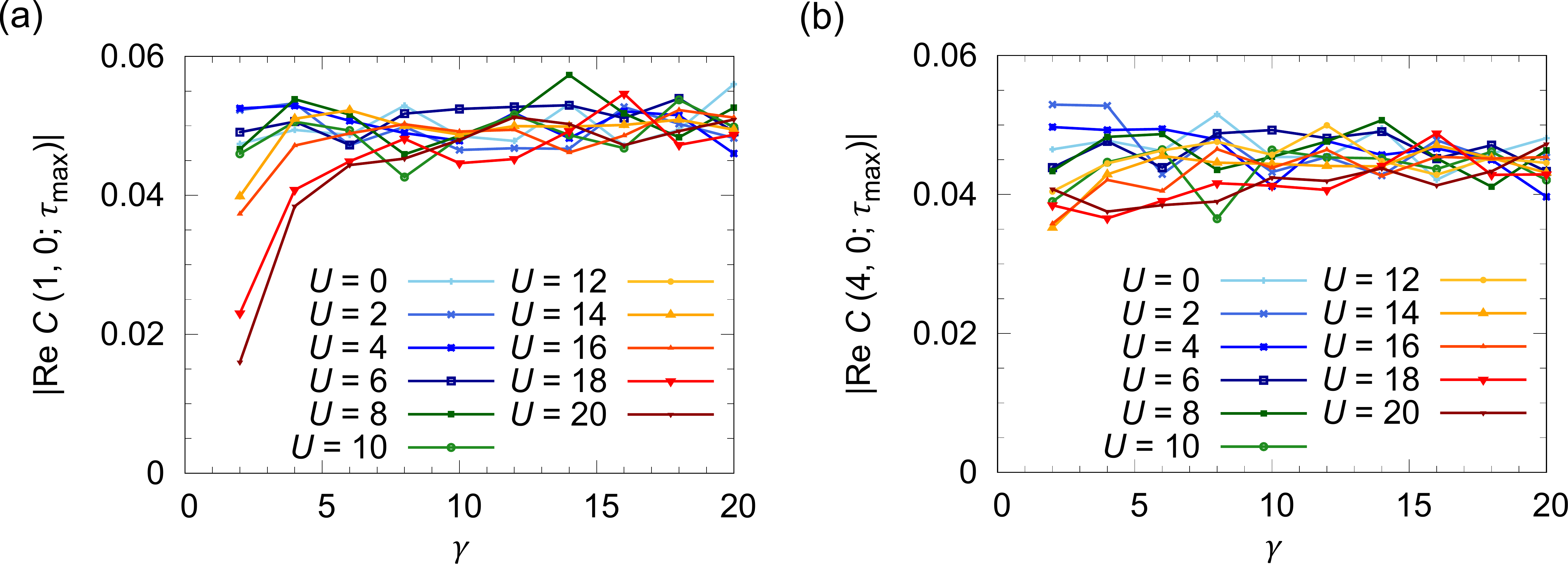}
\caption{Dependence of the pair correlations on the interaction strength $U$ and the spontaneous-emission rate $\gamma$. The initial state is the doublon-density-wave state of an 8-site system as in Fig.~\ref{fig_paircorr}(a) in the main text. The unit of energy is set to $t=1$. (a) Pair correlation $|\mathrm{Re}[C(1,0,\tau_{\mathrm{max}})]|$ between sites $j=1$ and $j=0$ at time $\tau_{\mathrm{max}}/\tau_h=40$. (b) Pair correlation $|\mathrm{Re}[C(4,0,\tau_{\mathrm{max}})]|$ between sites $j=4$ and $j=0$ at time $\tau_{\mathrm{max}}/\tau_h=40$.}
\label{fig_parameter}
\end{figure}

In Fig.~\ref{fig_parameter}, we show the dependence of the pair correlation function on the model parameters $U$ and $\gamma$. Here, we take an initial state $\rho(0)=\ket{\psi_0}\bra{\psi_0}$ of an 8-site system, where $\ket{\psi_0}=\prod_{j=0,2,4,6}c_{j\up}^\dag c_{j\down}^\dag\ket{0}$, as in Fig.~\ref{fig_paircorr}(a) in the main text. The pair correlation $C(1,0;\tau_{\mathrm{max}})$ ($C(4,0;\tau_{\mathrm{max}})$) between the zeroth site and the first (fourth) site is evaluated at $\tau_{\mathrm{max}}/\tau_h=40$. As shown in Fig.~\ref{fig_parameter}, the parameter dependence of the pair correlation is weak. The decrease in pair correlations for large $U$ and small $\gamma$ in Fig.~\ref{fig_parameter}(a) is attributed to the fact that the system does not reach a steady state at $\tau_{\mathrm{max}}/\tau_h=40$ due to the slow relaxation in this parameter regime where $|\mathrm{Im}[J]|\simeq 0.01$. This result indicates that the $\eta$-pairing state can be realized over a wide range of parameters of the model, if the initial state is appropriately chosen.

\section{Effect of dephasing}

\begin{figure}
\includegraphics[width=17.5cm]{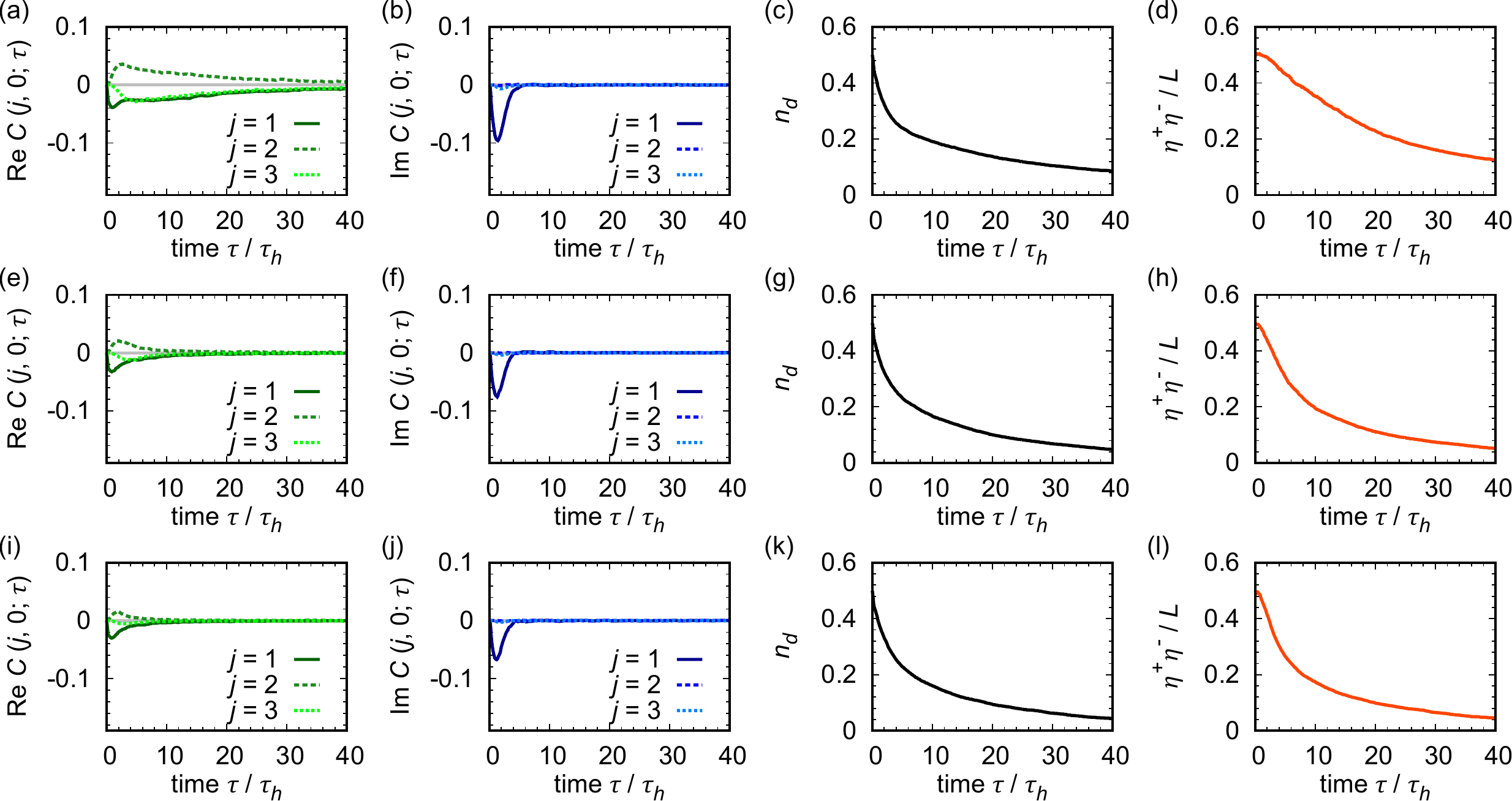}
\caption{Dynamics of $\eta$-pairing formation in the presence of dephasing. (a) (e) (i) Real part of the pair correlation functions $C(j,0,\tau)$. (b) (f) (j) Imaginary part of the pair correlation functions. (c) (g) (k) Density of doublons. (d) (h) (l) $\langle\eta^+\eta^-\rangle/L$ defined in Eq.~\eqref{eq_conserv} in the main text. The initial state is the same as in Figs.~\ref{fig_paircorr} (a) and (b). The model parameters are set to $t=1,U=10,\delta=0$, and $\gamma=10$, and the rate of dephasing is $\gamma_d=0.1$ for (a)-(d), $\gamma_d=0.5$ for (e)-(h), and $\gamma_d=1$ for (i)-(l). The unit of time is $\tau_h=1/t$.}
\label{fig_deph}
\end{figure}

As described in the experimental implementation of the model, the second-order process involving spontaneous decay from the $\ket{e}$ state to the $\ket{g,\up}$ state is described by an additional Lindblad operator $L_j^{(d)}$ which leads to dephasing of spin-up particles. In Fig.~\ref{fig_deph}, we show the effect of dephasing on $\eta$-pairing formation by numerically solving the quantum master equation \eqref{eq_master_deph} with the quantum-trajectory method. The initial state and the model parameters except for $\gamma_{d}$ are the same as those in Figs.~\ref{fig_paircorr} (a) and (b). Since the $\eta$-pairing states are not steady states in the presence of nonzero $\gamma_d$, the pair correlations decrease in a time scale which is roughly given by $1/\gamma_d$ [see Figs.~\ref{fig_deph} (a), (e), and (i)]. The imaginary part of each pair correlation function only shows a transient increase and it is negligible after a long time for all parameter regimes including the case of $\gamma_d=0$ [see Figs.~\ref{fig_deph} (b), (f), and (j)]. While the density of doublons stays nonvanishing as shown in Fig.~\ref{fig_paircorr}(b), it slowly decreases in the presence of dephasing [see Figs.~\ref{fig_deph} (c), (g), and (k)], implying that the steady state under nonzero $\gamma_d$ is a fully polarized state that does not contain spin-up particles. This behavior is consistent with the fact that the quantity $C=\langle\eta^+\eta^-\rangle$ [see Eq.~\eqref{eq_conserv} in the main text] is not conserved in the presence of nonzero $\gamma_d$, as shown in Figs.~\ref{fig_deph} (d), (h), and (l). Thus, if the condition $\gamma_d\ll\gamma,t$ is satisfied, the formation of the $\eta$-pairing state can be observed during its lifetime determined by $1/\gamma_d$.

\begin{figure}
\includegraphics[width=17.5cm]{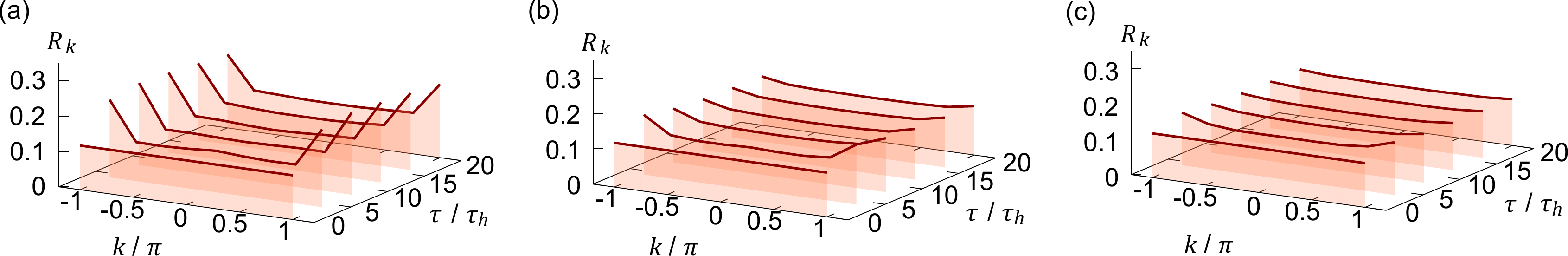}
\caption{Dynamics of the normalized momentum distribution $R_k$ of doublons in the presence of dephasing. The initial state and the model parameters $t,U,\delta,\gamma$ are the same as in Fig.~\ref{fig_momdist}(a), and the rate of dephasing is (a) $\gamma_d=0.1$, (b) $\gamma_d=0.5$, and (c) $\gamma_d=1$.}
\label{fig_deph_Rdist}
\end{figure}

In Fig.~\ref{fig_deph_Rdist}, we also show the momentum distribution of doublons in the presence of dephasing. Although the dephasing suppresses the peaks at $k=\pm \pi$, one can still observe the signature of $\eta$-pairing formation unless the rate of dephasing is large.

\end{document}